\documentclass{article}

\usepackage[dvipsnames]{xcolor}
\usepackage{color, colortbl}
\usepackage{float}
\usepackage{microtype}
\usepackage{graphicx}
\usepackage[export]{adjustbox}
\usepackage{subcaption}
\usepackage{booktabs}
\usepackage{tabularx}
\usepackage{multicol}
\usepackage{multirow}
\usepackage{pgfplots,gincltex}
\usepgfplotslibrary{groupplots}
\usepackage{tikz}
\usetikzlibrary{arrows.meta}

\pgfplotsset{width=\columnwidth,compat=newest,
       /pgfplots/ybar legend/.style={
        /pgfplots/legend image code/.code={%
        \draw[##1,/tikz/.cd,bar width=3pt,yshift=-0.2em,bar shift=0pt]
                plot coordinates {(0cm,0.8em)};},}}

\usepackage{hyperref}

\newcolumntype{C}{>{\centering\arraybackslash}X}

\usepackage[accepted]{icml2024}

\usepackage{amsmath}
\usepackage{amssymb}
\usepackage{mathtools}
\usepackage{amsthm}
\usepackage[capitalize,noabbrev]{cleveref}

\theoremstyle{plain}

\theoremstyle{definition}

\theoremstyle{remark}

\definecolor{cvprblue}{rgb}{0.21,0.49,0.74}
\definecolor{pltblue}{RGB}{174, 199, 232}
\definecolor{pltorange}{RGB}{255, 229, 204}
\definecolor{pltgreen}{RGB}{204, 229, 204}
\definecolor{pltred}{RGB}{229, 204, 204}
\definecolor{pltpurple}{RGB}{239, 218, 230}
\definecolor{cblue}{RGB}{52, 104, 192}
\definecolor{lblue}{RGB}{134, 167, 252}
\definecolor{corange}{RGB}{255, 152, 67}
\definecolor{lorgange}{RGB}{255, 221, 149}

\usepackage[textsize=tiny]{todonotes}

\icmltitlerunning{See More Details: Efficient Image Super-Resolution by Experts Mining}

\begin{document}

\twocolumn[
\icmltitle{See More Details: Efficient Image Super-Resolution by Experts Mining}



\icmlsetsymbol{equal}{*}
\icmlsetsymbol{corresponding}{*}

\begin{icmlauthorlist}
\icmlauthor{Eduard Zamfir}{yyy}
\icmlauthor{Zongwei Wu}{yyy,corresponding}
\icmlauthor{Nancy Mehta}{yyy}
\icmlauthor{Yulun Zhang}{sjt,sch,corresponding}
\icmlauthor{Radu Timofte}{yyy}

\end{icmlauthorlist}

\icmlaffiliation{yyy}{Computer Vision Lab, CAIDAS \& IFI, University of Würzburg, Germany}
\icmlaffiliation{sjt}{AI Institute, Shanghai Jiao Tong University, China}
\icmlaffiliation{sch}{Computer Vision Lab, ETH Zurich, Switzerland}

\icmlcorrespondingauthor{Zongwei Wu}{zongwei.wu@uni-wuerzburg.de}
\icmlcorrespondingauthor{Yulun Zhang}{yulun100@gmail.com}

\icmlkeywords{Mixture of Experts, Efficiency, Image Super-Resolution}

\vskip 0.3in
]



\printAffiliationsAndNotice{}  

\begin{abstract}
Reconstructing high-resolution (HR) images from low-resolution (LR) inputs poses a significant challenge in image super-resolution (SR). While recent approaches have demonstrated the efficacy of intricate operations customized for various objectives, the straightforward stacking of these disparate operations can result in a substantial computational burden, hampering their practical utility. In response, we introduce \textbf{S}eemo\textbf{R}e, an efficient SR model employing expert mining. Our approach strategically incorporates experts at different levels, adopting a collaborative methodology. At the macro scale, our experts address rank-wise and spatial-wise informative features, providing a holistic understanding. Subsequently, the model delves into the subtleties of rank choice by leveraging a mixture of low-rank experts. By tapping into experts specialized in distinct key factors crucial for accurate SR, our model excels in uncovering intricate intra-feature details. This collaborative approach is reminiscent of the concept of ``see more", allowing our model to achieve an optimal performance with minimal computational costs in efficient settings. The source codes will be publicly made available at \url{https://github.com/eduardzamfir/seemoredetails}

\end{abstract}

\section{Introduction}
\label{sec:intro}
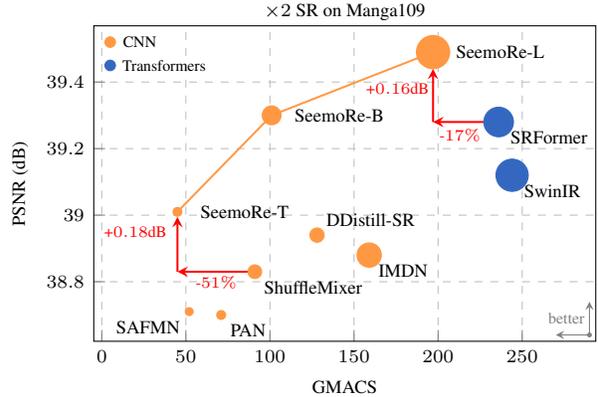
\begin{figure}[t]
    \centering
        \begin{tikzpicture}[baseline]

        \begin{axis}[
            width=\columnwidth,
            height=0.7\columnwidth,
            xtick pos=left,
            ytick pos=left,
            xlabel near ticks,
            ylabel near ticks,  
            xlabel={GMACS},
            ylabel={PSNR (dB)},
            grid,
            grid style=dashed,
            title={$\times 2$ SR on Manga109},
            title style={anchor=north, at={(0.5,1.05)},},
            tick label style={font=\scriptsize}, 
            title style={font=\scriptsize}, 
            label style={font=\scriptsize}, 
            nodes near coords, 
            enlarge x limits=0.25,
            enlarge y limits=0.1,
            legend style={at={(.2,.925)}, anchor=center, legend columns=-1, font=\tiny,},
        ]
    \addplot[
        nodes near coords style = {font=\scriptsize, anchor=north west},
        scatter/classes={
                b={color=corange, mark size=1.7},
                c={color=corange, mark size=1.5},
                d={color=corange, mark size=2.6},
                f={color=corange, mark size=4.6},
                b={color=corange, mark size=1.7}
                },
        scatter, mark=*, only marks, 
        scatter src=explicit symbolic,
        nodes near coords*={\Label},
        visualization depends on={value \thisrow{label} \as \Label} 
        ] table [meta=class] {
            x y params class label
            91 38.83 394 d ShuffleMixer
            159 38.88 694 f IMDN
           71 38.70 261 b PAN 
        };

      \addplot[
        nodes near coords style = {font=\scriptsize, anchor=south west},
        scatter/classes={
                h={color=corange, mark size=2.7}
                },
        scatter, mark=*, only marks, 
        scatter src=explicit symbolic,
        nodes near coords*={\Label},
        visualization depends on={value \thisrow{label} \as \Label} 
        ] table [meta=class] {
            x y params class label
            128 38.94 414 h DDistill-SR
        };

     \addplot[
        nodes near coords style = {font=\scriptsize, anchor=north east},
        scatter/classes={
                c={color=corange, mark size=1.5}
                },
        scatter, mark=*, only marks, 
        scatter src=explicit symbolic,
        nodes near coords*={\Label},
        visualization depends on={value \thisrow{label} \as \Label} 
        ] table [meta=class] {
            x y params class label
            52 38.71 228 c SAFMN
            
        };
    
    \addplot[
        scatter/classes={
                a={mark size=1.5},
                b={mark size=3.4},
                c={mark size=6.12}},
        scatter, mark=*, only marks, sharp plot,corange, line width=0.75pt,
        scatter src=explicit symbolic,
        nodes near coords style={text=black, font=\scriptsize, anchor=south west, yshift=-6pt, xshift=5pt},
        nodes near coords*={\Label},
        visualization depends on={value \thisrow{label} \as \Label} 
        ] table [meta=class] {
        x y label class
        45 39.01  SeemoRe-T a 
        101 39.30  SeemoRe-B b
        197 39.49 SeemoRe-L c
        };

    \addplot[
        nodes near coords style = {font=\scriptsize, anchor=north west,xshift=1pt},
        scatter/classes={
            h={mark size=6.1,},
            j={mark size=5.6,}
            },
        scatter, mark=*, only marks, cblue, 
        scatter src=explicit symbolic,
        nodes near coords style={text=black},
        nodes near coords*={\Label},
        visualization depends on={value \thisrow{label} \as \Label}
        ] table [meta=class] {
                    x y params class label
                    236 39.28 853 j SRFormer
                    244 39.12  910 h SwinIR
        };


    \draw[-stealth, red, line width=0.75pt] (axis cs:45, 38.83) -- (axis cs:45, 38.995);
    \draw[-stealth, red, line width=0.75pt] (axis cs: 91, 38.83 ) -- (axis cs: 45, 38.83 );
    \node at (axis cs:20,38.95) {\tiny\textcolor{red}{+$0.18$dB}};
    \node at (axis cs:68,38.80) {\tiny\textcolor{red}{-$51\%$}};

    \draw[-stealth, red, line width=0.75pt] (axis cs:197, 39.28) -- (axis cs:197, 39.44);
    \draw[-stealth, red, line width=0.75pt] (axis cs: 236, 39.28) -- (axis cs: 197, 39.28);
    \node at (axis cs:176,39.38) {\tiny\textcolor{red}{+$0.16$dB}};
    \node at (axis cs:213,39.24) {\tiny\textcolor{red}{-$17\%$}};

    \draw[draw=gray, fill=gray] (290,38.64) circle [radius=0.025cm];
    \draw[-stealth,draw=gray,] (290,38.64) -- (290,38.74);
    \draw[-stealth,draw=gray] (290,38.64) -- (270,38.64);
    \node[text width=1cm] at (axis cs:288,38.69) {\tiny\textcolor{gray}{better}};

    \draw[draw=corange, fill=corange] (5,39.52) circle [radius=0.07cm];
    \node[text width=1cm] at (35,39.52) {\tiny CNN};
    \draw[draw=cblue, fill=cblue] (5,39.45) circle [radius=0.07cm];
    \node[text width=1cm] at (35,39.45) {\tiny Transformers};

        \end{axis}
    \end{tikzpicture}
    \caption{\textit{Model complexity trade-off.}  Visualization of PSNR, GMACS, and parameter counts on Manga109 dataset for $\times$2 task. Our proposed SeemoRe excels the state-of-the-art CNN-based and \textit{lightweight} Transformer-based SR models. Marker size indicates parameter counts w.r.t SwinIR-Light~\cite{liu2021swin}.}
    \label{fig:exp:efficiency}
\end{figure}
    Single image super-resolution (SR) is a long-standing low-level vision endeavour that pursues the reconstruction of a high-resolution (HR) image from its degraded low-resolution (LR) counterpart.
    This challenging task has garnered considerable attention owing to the expeditious development of ultra-high definition devices and video streaming applications ~\cite{khani2021efficient,zhang2021benchmarking}. 
Foreseeing the resource constraints, it is of substantial desire to design an efficient SR model for gauging the HR images to be perfectly visualized on these devices or platforms.
Identifying the most plausible candidates for missing HR pixels poses a particular challenge for SR. In the absence of external priors, the primary approaches for SR involves exploring the intricate relationships among the neighboring pixels for reconstruction. Recent SR models exemplify this through methods such as \textbf{(a)} attention ~\cite{liang2021swinir,zhou2023srformer,chen2023dat}, \textbf{(b)} feature mixing ~\cite{hou2022conv2former,sun2023safmn}, and \textbf{(c)} global-local context modeling ~\cite{wang2023omnisr,sun2022shufflemixer}, yielding remarkable accuracy.

Unlike other approaches in this work, we aim to avoid complex and disconnected blocks focusing on specific factors, opting instead for a unified learning module specialized for all aspects. However, an additional challenge arises due to the efficiency requirement, rendering implicit learning through a vast number of parameters unfeasible, especially in the context of devices with limited resources.

To achieve such an efficient unification, we introduce \textbf{S}eemo\textbf{R}e, which leverages the synergy of different experts to maximize intra-feature intertwining, collaboratively learning a cohesive relation across LR pixels. Our motivation stems from the observation that image features often display diverse patterns and structures. Attempting to capture and model all these patterns with a single, monolithic model can be challenging. Collaborative experts, on the other hand, enable the network to specialize in different regions or aspects of the input space, enhancing its adaptability to various patterns and facilitating the modeling of LR-HR dependencies, akin to ``See More".

Technically, our network is composed of stacked residual groups (RGs) for dynamically selecting the pivotal features via experts, focusing on two different aspects. 
At the macro level, each RG embodies two successive expert blocks: \textbf{(a)} \textit{Rank modulating expert} (RME), expertized in dealing with the most informative features through low-rank modulation, and \textbf{(b)} \textit{Spatial modulating expert} (SME), expertized in efficient spatial enhancement. 
At the micro level, we devise a Mixture of Low-Rank Expertise (MoRE) as the foundational component within RME to dynamically select the best and most suitable rank for different inputs and at different network depths while implicitly modeling the global contextual relationships. 
Furthermore, we design a Spatial Enhancement Expertise (SEE) as an efficient alternative to complex self-attention within SME for distinctly improving the spatial-wise local aggregation capabilities. 
Such a combination efficiently modulates the mutual dependencies within the feature attributes, enabling our model to extract high-level information, which is a key aspect of SR. By explicitly mining experts at different granularity for different expertise, our network navigates the intricacies between spatial and channel features, maximizing their synergistic contribution and thus accurately and efficiently reconstructing more details. 

As shown in Figure \ref{fig:exp:efficiency}, our network significantly outperforms the state-of-the-art (SOTA) efficient models such as DDistill-SR~\cite{wang2022ddistill} or SAFMN~\cite{sun2023safmn} by a considerable margin, while utilizing only half or even less of the GMACS. Although our model is specifically designed for efficient SR, its scalability is evident as our larger model surpasses the SOTA lightweight transformer in performance while incurring lower computational costs. Overall, our key contributions are threefold:
    \begin{itemize} 
       \item We propose SeemoRe which matches the versatility of Transformer-based methods and the efficiency of CNN-based methods. 
       \item A Rank modulating expert (RME) is proposed to probe into the intricate inter-dependencies among the relevant feature projections in an efficient manner. 
       \item A Spatial modulating expert (SME) is proposed to integrate the complementary features extracted by SME by encoding the local contextual information. 
    \end{itemize}

\section{Related Works}
\label{sec:related}
\paragraph{CNN-based SR.} 
In recent years, CNN-based techniques have outperformed traditional interpolation algorithms~\cite{LanczosFilteringinOneandTwoDimensions} by learning a non-linear mapping between the input and target in an end-to-end training manner. The seminal SRCNN~\cite{dong2014learning} introduced a three-layer convolutional approach for image super-resolution, later extended by works such as~\cite{lim2017enhanced,zhang2018residual,hui2019imdn,liang2021swinir}. VDSR~\cite{kim2016deeply} and EDSR~\cite{lim2017enhanced} deepen networks using residual learning principles, with EDSR streamlining residual blocks for deeper training. Conversely, RCAN~\cite{zhang2018rcan} introduces a novel residual-in-residual architecture for models exceeding 400 layers. While various spatial and channel attention mechanisms aim to enhance image reconstruction quality, CNN-based techniques still struggle to effectively utilize shared information across both dimensions. In this work, we aim to explore the interdependencies among the features in a computationally efficient way.

\paragraph{Transformer-based SR.}
Thanks to its remarkable performance in high-level tasks~\cite{dosovitskiy2021an}, the Transformer architecture has found its way into low-level vision tasks, such as image SR. Contemporary Transformer-based approaches aim to alleviate the computational load by confining self-attention to local regions and incorporating a higher degree of locality bias into their network design. SwinIR~\cite{liang2021swinir} incorporates local window self-attention and a shift mechanism inspired by the Swin Transformer design\cite{liu2021swin}. Meanwhile, others like ELAN \cite{zhang2022elan} or ESRT ~\cite{lu2022esrt} reduce the feature dimensions by splitting or down-scaling to enhance the computational efficiency. Omni-SR~\cite{wang2023omnisr} models pixel-interactions across different axes, creating universal correlations. SRFormer~\cite{zhou2023srformer} optimizes the computational efficiency by employing large window self-attention through the permutation of self-attention mechanisms. However, transformer-based methods typically demand significantly higher computational resources, even with smaller model capacities.

\begin{figure*}[t]
    \centering
    \includegraphics[width=\textwidth]{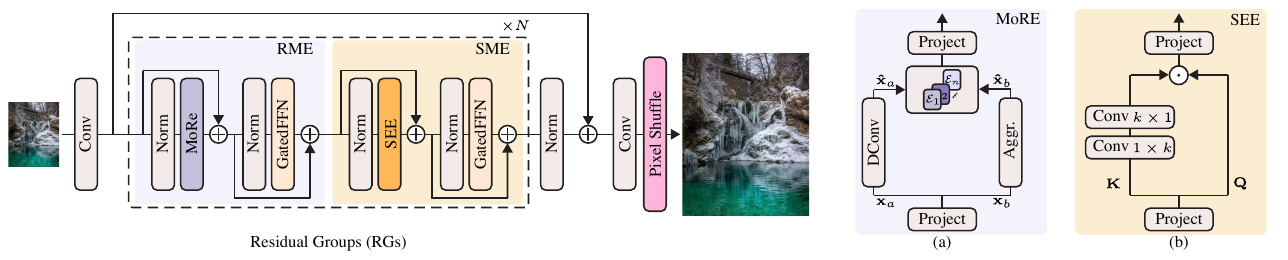}
    \vspace{-6mm}
    \caption{\textit{Architecture Overview}. \textbf{S}eemo\textbf{R}e refines the feature representations via stacked Residual groups (RGs). Each RG consists of a Rank Modulating Exert (RME) and a Spatial Modulating Expert (SME). RME leverages the Mixture of Low Rank Expertise (MoRE) to refine the global texture, while SME employs spatial enhancement experts (SEE) to supplement RME with spatial cues.}
    \label{fig:main_arch}
\end{figure*}

\paragraph{Efficiency in SR.}
In recent years, the pursuit of efficient SR techniques has gained significant momentum~\cite{li2022ntire, ignatov2023efficient, li2023ntire_esr, conde2023rtsr}. Consequently, researchers have introduced streamlined neural architectures~\cite{ignatov2021real}, network compression~\cite{wang2022ddistill}, reparameterization~\cite{zhang2021ecbsr}, and other training strategies to cater to the demand for efficiency. 
Initially, efficient SR methods utilized group convolutions and cascaded block designs to boost efficiency~\cite{ahn2018carn,hui2019imdn}. Subsequent advancements introduced convolution-based spatial or channel enhancement modules~\cite{liu2020residual}. More recently, ShuffleMixer~\cite{sun2022shufflemixer} integrates large kernel convolutions and feature shuffling, improving both computational efficiency and high-resolution reconstruction. SAFMN~\cite{sun2023safmn} improves the efficiency by collecting non-local features using a shallow pyramid. Despite  improvements in several efficiency aspects brought up by the aforementioned approaches, there is still scope for a better trade-off between model efficiency and the restoration performance.

\paragraph{Dynamic Networks.}
Dynamic networks have been extensively studied to optimize the balance between speed and performance across various tasks. Early research employed conditional computation to selectively activate network segments at different times~\cite{bengio2013estimating}. More recently, Mixture-of-Experts (MoE) approaches with routing architecture~\cite{shazeer2017moe, riquelme2021scaling, puigcerver2024softmoe} have expanded model capacity without significantly increasing inference costs, primarily enhancing the feed-forward capacity of Transformers in Natural language processing~\cite{shazeer2017moe} and high-level vision tasks~\cite{riquelme2021scaling, puigcerver2024softmoe}. A similar idea can be found in image restoration, where Path-Restore~\cite{yu2021path} dynamically routes image patches to different network paths based on content and distortion, leveraging a difficulty-regulated reward function. In this work, our research explores the routing concept from an architecture design perspective for image super-resolution, aiming to discover the most efficient and appropriate expert to improve the feature modeling. 

\section{Methodology}

In this section, we unveil the fundamental components of our proposed model tailored for efficient super-resolution. As demonstrated in \cref{fig:main_arch}, our overall pipeline embodies a sequence of $N$ residual groups (RGs) and an upsampler layer. The initial step involves applying a 3$\times$3 convolution layer to generate the shallow features from the input low-resolution (LR) image. Subsequently, multiple stacked RGs are deployed to refine the deep features, easing the reconstruction of high-resolution (HR) images while maintaining efficiency. Each RG consists of a Rank modulating expert (RME) and a Spatial modulating expert (SME). Lastly, a global residual connection links the shallow features to the output of the deep features for capturing the high-frequency details and an up-sampler layer ( 3$\times$3 and pixel-shuffle~\cite{shi2016real}) is deployed for faster reconstruction.

\subsection{Rank Modulating Expert}

Unlike large kernel convolution \cite{hou2022conv2former} or self-attention \cite{vaswani2017attention} that rely upon resource-intensive matrix operations for modelling the LR-HR dependencies, we opt for modulating the most relevant interactions in low-rank in our quest for efficiency. 
Our proposed Rank modulating expert (RME) (see \cref{fig:main_arch}) explores a Transformer alike architecture using Mixture of Low-Rank Expertise (MoRE) for modelling the relevant global informative features efficiently and a GatedFFN \cite{chen2023dat} for refined contextual feature aggregation.

\begin{figure}[t]
    \includegraphics[width=\columnwidth]{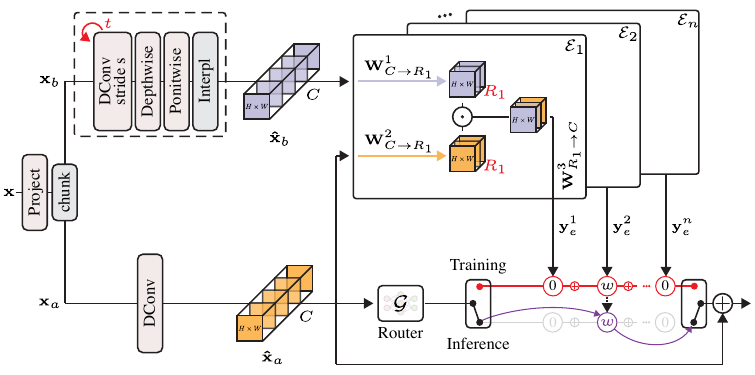}
     \caption{Illustration of the proposed Mixture of Low-Rank Expertise (MoRE) as a core block of the RME.}
    \label{fig:more}
\end{figure}

\paragraph{Mixture of Low-Rank Expertise.} As illustrated in \cref{fig:more}, from a layer normalised input tensor $\mathbf{x} \in \mathbb{R}^{H \times W \times C}$, we use a 3$\times$3 convolution for feature projection and then we split along the channel dimension to create two distinct views $\mathbf{x}_{a}~\text{and}~\mathbf{x}_{b} \in \mathbb{R}^{H \times W \times C}$. To efficiently aggregate the pixel-wise cross-channel context, we leverage a recursive strided convolution $t$ times followed by a refinement and upsampling step, resulting in the construction of the feature pyramid denoted as $\mathbf{\hat{x}}_{b} \in \mathbb{R}^{H \times W \times C}$. The process is formulated as follows:
\begin{align}
    \label{eq:agg}
     |p|_{\downarrow h \times w}  &= \text{DConv}^{s}_{k \times k}(...(\text{DConv}^{s}_{k \times k}(\mathbf{x}_{b}))\\
     \mathbf{\hat{x}}_{b} &= |~\mathbf{W}_{C \rightarrow C}(\text{DConv}_{3 \times 3}(|p|_{\downarrow h \times w}))~|_{\uparrow H \times W}, 
\end{align}
where $\text{DConv}_{k \times k}$ denotes a depth-wise convolution with kernel size $k$ and stride $s$, $\mathbf{W}_{C \rightarrow C}$ denotes a linear layer, $p$ represents the contextual feature pyramid. Simultaneously, a parallel depth-wise convolution extracts the local spatial context $\mathbf{\hat{x}}_{a}$ before feeding both the extracted feature maps into the mixture of low-rank expertise. This branched parallel design approach is chosen purposefully. In general, the downsampling of the feature maps impacts the reconstruction performance of SR methods. Therefore, we maintain the same resolution for general feature extraction while incorporating an additional path to capture global contextual cues efficiently, thereby circumventing any information loss.

To further delve into the intricacies of the inter-dependencies among the extracted features for reducing complexity, we deploy low-rank decomposition for the inputs while modeling the global contextual relationships. As demonstrated in \cref{fig:more}, a single low-rank expert ($\mathcal{E}$), takes as input the spatial features, $\mathbf{\hat{x}}_{a}$, and encoded pixel-wise contextual cues, $\mathbf{\hat{x}}_{b}$ and is formulated as:\begin{align}
    \label{eq:expert}
    \mathcal{E}_i &= \mathbf{W}^{3}_{R_{i} \rightarrow C} (\mathbf{W}^{1}_{C \rightarrow R_{i}}\mathbf{\hat{x}}_{a} \odot \mathbf{W}^{2}_{C \rightarrow R_{i}}\mathbf{\hat{x}}_{b}),
\end{align}
where the linear layers  denoted as $\mathbf{W}_{C \rightarrow R_{i}}$, compress the encoded features along the channel dimension to their low-rank approximation $R_{i}$, where $i \in \{1,...,n\}$.  After adeptly modulating the spatial cues through element-wise multiplication with the contextual cues in low-dimensional space, another linear layer $\mathbf{W}^{3}_{R_{i} \rightarrow C}$ extends the features back to the original dimension $C$ to extract the relevant channel-wise spatial content. Thereby, implicitly mixing the crucial spatial and channel dependencies in an efficient way. 

\begin{algorithm}[t]
    \caption{\footnotesize Mixture of Low-Rank Experise}
    \label{alg:moe_layer}
\begin{algorithmic}[1]
    \STATE \textbf{Input:} Input feature $\mathbf{\hat{x}}_{a}$, semantic cues $\mathbf{\hat{x}}_{b}$
    \STATE \textbf{Parameters:} $n$ Experts $\mathcal{E}$, Router $\mathcal{G}$, Low-Rank dimensions $R_{i} = 2^{i+1}$ with $i \in \{1,...,n\}$
    , top-1 expert $k=1$
    
    \STATE Compute router outputs: $\mathbf{g} = \mathcal{G}(\mathbf{\hat{x}}_{a})$
    \STATE Normalize weights: $\mathbf{w} = \text{Softmax}(\mathbf{g})$
    \STATE Select top-1 expert: $w_{\text{top-1}} = \text{topk}(\mathbf{w}, k=1)$
    \STATE Set all other weights to zero: $\mathbf{w}_i = 0$ for $i \neq \text{top-1}$
    
    \IF{training}
        \FOR{each $e \in \mathcal{E}$}
            \STATE $\mathbf{y}^{i}_{e} = \mathbf{W}^{3}_{R_{i} \rightarrow C} (\mathbf{W}^{1}_{C \rightarrow R_{i}}\mathbf{\hat{x}}_{a} \odot \mathbf{W}^{2}_{C \rightarrow R_{i}}\mathbf{\hat{x}}_{b})$
        \ENDFOR
        \STATE Compute final output: $\mathbf{y} = \sum_{i=1}^{n} w_i \cdot \mathbf{y}^{i}_{e}$
    \ELSE
        \STATE Compute final output: $\mathbf{y} = w_{\text{top-1}} \cdot \mathbf{y}^{\text{top-1}}_{e}$
    \ENDIF
    
    \STATE \textbf{Output:} Final output $\mathbf{y}$
\end{algorithmic}
\end{algorithm}

However, manually determining the optimal low-rank ($R$) may not fully leverage all the inherent information for modulation, leading to underutilized model capacity. Thus, we employ a dynamic approach using a mixture of different low rank experts, with a routing network ($\mathcal{G}$) that systematically explores the search space to identify the ideal low-rank expert based on the input and network depth. Following \cite{shazeer2017moe}, the final output $\mathbf{y}$ of the mixture of low-rank experts is as follows:
\begin{align}
    \mathbf{y} &= \sum_{i}^{n} \mathcal{G}(\mathbf{\hat{x}}_{a})\mathcal{E}_i(\mathbf{\hat{x}}_{a},\mathbf{\hat{x}}_{b}) + \mathbf{\hat{x}}_{a},
\end{align}
where $\mathcal{G}(\cdot)$ and $\mathcal{E}_i(\cdot)$ denote the learned routing function and the output of the $i$-th expert, respectively. The sparsity inherent in the router function $\mathcal{G}(\cdot)$ optimizes computation by assigning greater weights to the top-$k$ low-rank experts. While at training time, our method learns from different experts, during inference, only the selected top-$k$ expert is utilized for computation, further enhancing the efficiency. More specifically, the inference complexity is not proportional to the number of experts. 

Adhering to the MoE concept with \(k > 1\), our routing function for optimal low-rank representation extends sparse routing principles~\cite{shazeer2017moe} by selecting only the top-$1$ expert. As our work is pioneering in this domain, we emphasize a more interpretable top-1 design, as shown in \cref{fig:exp:lowrank_vis}, which allows us to streamline the model architecture and computational process, creating an efficient yet powerful image super-resolution model. 
Technically, both training and inference leverage dynamic expert selection based on input and model depth; however, only the top-$1$ expert per layer is utilized, with contributions from other experts weighted at zero. 
During inference, inactive experts are disregarded to efficiently exploit contextual information using the optimal input-dependent expert chosen by the router. 
This ensures consistency between training and inference, as only one expert per layer remains active, thereby mitigating potential discrepancies. In \cref{tab:supp:top_k} found in the supplementary, we show that augmenting the number of top-$k$ experts can slightly improve the performance, at the cost of increased computational complexity. We hope that our network can serve as a fresh baseline for future development.


Additionally, the design choices contributing towards the selection of the number of low-rank experts ($\mathcal{E}_i$) and the rank dimension ($R$) for memory-efficient reconstruction is illustrated in \cref{tab:exp:more_ablation} of the ablation study. We also provide the pseudocode for the proposed MoRE block in \cref{alg:moe_layer}.
In addition to the primary analyses presented in the main text, the supplementary material offer further insights and experiments that substantiate the design decisions of our proposed MoRE module. For detailed information, refer to \cref{tab:supp:more_design,tab:supp:num_exp}.

\subsection{Spatial Modulating Expert}
We observe that the rank modulating expert is more dedicated towards investigating the global channel-wise contextual information, and its effectiveness would be complemented by the spatial-wise local information. Inspired by the previous work in classification~\cite{yang2022focalnet,hou2022conv2former}, we design a spatial modulating expert (SME) (see \cref{fig:main_arch}) comprising of a spatial enhancement expertise (SEE) block that efficiently captures the spatial-wise coupling followed by a GatedFFN \cite{chen2023dat} for feature refinement.

\paragraph{Spatial Enhancement Expertise.} While the vanilla self-attention (SA) mechanism~\cite{vaswani2017attention} creates connections among all the input pixels, effectively capturing the relevant context, its quadratic computational complexity with image size poses limitations, particularly in high-resolution scenarios like image SR. Thus, our spatial enhancement expertise simplifies the computation of the similarity matrix $\mathbf{A}$ between keys $\mathbf{K}$ and queries $\mathbf{Q}$ by utilizing a striped depth-wise convolution with a large kernel, sequentially convolving the feature maps with $\mathbf{k_1} \in \mathbb{R}^{[1, k]}$ followed by $\mathbf{k_2} \in \mathbb{R}^{[k, 1]}$. Specifically, we compute the locally enhanced spatial-wise features as follows:
\begin{align}
    \label{eq:attn_block}
    \mathbf{x}_{out} &= \text{DConv}^{s}_{k \times k}(\mathbf{W}^{4}_{C \rightarrow C}\mathbf{x}_{in}) \odot \mathbf{W}^{5}_{C \rightarrow C}\mathbf{x}_{in},
\end{align}
where $\odot$ is the Hadamard product, $\mathbf{W}^{4}_{C \rightarrow C}$ and $\mathbf{W}^{5}_{C \rightarrow C}$ are linear (project) layers, $\text{DConv}^{s}_{k \times k}$ denotes the striped depth-wise convolution, and $\mathbf{x}_{in}$ is the layer normalised output of the RME.
The use of a large-kernel convolution facilitates a localized correlation among the pixels within the $k \times k$ window, emulating the window-based SA layers frequently employed in image restoration~\cite{liu2021swin,Zamir2021Restormer,chen2023dat}, all the while preserving the efficiency benefits associated with convolutional layers as demonstrated in \cref{tab:exp:arch_components}.

\section{Experiments}
\label{sec:experiments}

\begin{table*}[ht]
    \centering
    \footnotesize
    \fboxsep0.8pt
    \setlength\tabcolsep{3pt}   
    \caption{\textit{Comparison to efficient SR models.}~PSNR (dB $\uparrow$) and SSIM ($\uparrow$) metrics are reported on the Y-channel. \colorbox{red!10}{\textcolor{red}{Best}} and \colorbox{blue!10}{\textcolor{blue}{second best}} performances are highlighted. GMACS (G) are computed by upscaling to a $1280\times720$ HR image. SeemoRe-T achieves state-of-the-art performance across all benchmarks with the lowest parameter count and computational demand.
    `-' represents unreported results.}
    \label{tab:exp:efficient_sota}
    \begin{tabularx}{\textwidth}{lX*{14}{c}}
    \toprule
    & \multirow{2}{*}{Method} & \multirow{2}{*}{Params} & \multirow{2}{*}{GMACS} &  
    \multicolumn{2}{c}{Set5} & \multicolumn{2}{c}{Set14} & \multicolumn{2}{c}{BSD100} & \multicolumn{2}{c}{Urban100} & \multicolumn{2}{c}{Manga109} \\
    \cmidrule(lr){5-6} \cmidrule(lr){7-8} \cmidrule(lr){9-10} \cmidrule(lr){11-12} \cmidrule(lr){13-14} 
    & & & &PSNR&SSIM &PSNR&SSIM &PSNR&SSIM &PSNR&SSIM &PSNR&SSIM\\
    \midrule
    
    \multirow{10}*{\rotatebox{90}{$\times 2$}}
    & Bicubic & - & - & 33.66 & .9299 & 30.24 & .8688 & 29.56 & .8431 & 26.88 & .8403 & 30.80 & .9339\\
    & CARN-M~\cite{ahn2018carn} & 412K & 91 & 37.53 & .9583 & 33.26 & .9141 & 31.92 & .8960 & 31.23 & .9193 & - & - \\
    & IMDN~\cite{hui2019imdn} & 694K &159&38.00&.9605&\cellcolor{blue!10}{\textcolor{blue}{33.63}}&.9177&\cellcolor{blue!10}{\textcolor{blue}{32.19}}&.8996&32.17&\cellcolor{blue!10}{\textcolor{blue}{.9283}}&38.88&\cellcolor{blue!10}{\textcolor{blue}{.9774}}\\
    & PAN~\cite{zhao2020pan} & 261K & 71 & 38.00 & .9605 & 33.59 & .9181 & 32.18 & .8997 & 32.01 & .9273 & 38.70 & .9773\\
    & DRSAN~\cite{park2021drsan} & 370K & 86 & 37.99 & .9606 & 33.57 & .9177 & 32.16 & .8999 & 32.10 &.9279 & - & - \\
    & DDistill-SR~\cite{wang2022ddistill} & 414K & 128 & \cellcolor{blue!10}{\textcolor{blue}{38.03}} & \cellcolor{blue!10}{\textcolor{blue}{.9606}} & 33.61 & \cellcolor{blue!10}{\textcolor{blue}{.9182}} & \cellcolor{blue!10}{\textcolor{blue}{32.19}} & \cellcolor{blue!10}{\textcolor{blue}{.9000}} & \cellcolor{blue!10}{\textcolor{blue}{32.18}} & \cellcolor{red!10}{\textcolor{red}{.9286}} & \cellcolor{blue!10}{\textcolor{blue}{38.94}} & \cellcolor{red!10}{\textcolor{red}{.9777}}\\
    & ShuffleMixer~\cite{sun2022shufflemixer} & 394K & 91 & 38.01 & \cellcolor{blue!10}{\textcolor{blue}{.9606}} & \cellcolor{blue!10}{\textcolor{blue}{33.63}} & .9180 & 32.17 & .8995 & 31.89 & .9257 & 38.83 & \cellcolor{blue!10}{\textcolor{blue}{.9774}}\\
    & SAFMN~\cite{sun2023safmn} & 228K & 52 & 38.00 & .9605 & 33.54 & .9177 & 32.16 & .8995 & 31.84 & .9256 & 38.71 & .9771\\
    & SeemoRe-T (\textit{ours})  & 220K & 45  & \cellcolor{red!10}{\textcolor{red}{38.06}} & \cellcolor{red!10}{\textcolor{red}{.9608}} & \cellcolor{red!10}{\textcolor{red}{33.65}} & \cellcolor{red!10}{\textcolor{red}{.9186}} & \cellcolor{red!10}{\textcolor{red}{32.23}} & \cellcolor{red!10}{\textcolor{red}{.9004}} & \cellcolor{red!10}{\textcolor{red}{32.22}} & \cellcolor{red!10}{\textcolor{red}{.9286}} & \cellcolor{red!10}{\textcolor{red}{39.01}} & \cellcolor{red!10}{\textcolor{red}{.9777}} \\

    \midrule

     \multirow{10}*{\rotatebox{90}{$\times 3$}}
    & Bicubic & - & - & 30.39&.8682&27.55&.7742&27.21&.7385&24.46&.7349&26.95&.8556\\
    & CARN-M~\cite{ahn2018carn} &  415K & 46 & 33.99&.9236 & 30.08 &.8367 & 28.91 & .8000 & 27.55 & .8385 & - & - \\
    & IMDN~\cite{hui2019imdn} & 703K &72 &34.36&.9270 & 30.32 & .8417 & 29.09 & .8046 & 28.17 & .8519 & 33.61 & .9445\\
    & PAN~\cite{zhao2020pan} & 261K & 39 & 34.40 & .9271 & 30.36 & \cellcolor{blue!10}{\textcolor{blue}{.8423}} & 29.11 & .8050 & 28.11 & .8511 & 33.61 & .9448\\
    & DRSAN~\cite{park2021drsan} & 410K & 43 & \cellcolor{blue!10}{\textcolor{blue}{34.41}} & .9272 & 30.27 & .8413 & 29.08 & \cellcolor{blue!10}{\textcolor{blue}{.8056}} & \cellcolor{blue!10}{\textcolor{blue}{28.19}} &\cellcolor{blue!10}{\textcolor{blue}{.8529}}& - & - \\
    & DDistill-SR~\cite{wang2022ddistill} & 414K & 57 & 34.37 & \cellcolor{blue!10}{\textcolor{blue}{.9275}} & 30.34 & .8420 & 29.11 & .8053 & \cellcolor{blue!10}{\textcolor{blue}{28.19}} & .8528 & \cellcolor{blue!10}{\textcolor{blue}{33.69}} & \cellcolor{blue!10}{\textcolor{blue}{.9451}} \\
    & ShuffleMixer~\cite{sun2022shufflemixer} & 415K & 42 & 34.40 & .9272 & \cellcolor{blue!10}{\textcolor{blue}{30.37}} & \cellcolor{blue!10}{\textcolor{blue}{.8423}} & \cellcolor{blue!10}{\textcolor{blue}{29.12}} & .8051 & 28.08 & .8498 & \cellcolor{blue!10}{\textcolor{blue}{33.69}} & .9448 \\
    & SAFMN~\cite{sun2023safmn} & 233K & 23 & 34.34 & .9267 & 30.33 & .8418 & 29.08 & .8048 & 27.95 & .8474 & 33.52 & .9437  \\
    & SeemoRe-T (\textit{ours}) & 225K & 20 &  \cellcolor{red!10}{\textcolor{red}{34.46}} & \cellcolor{red!10}{\textcolor{red}{.9276}} & \cellcolor{red!10}{\textcolor{red}{30.44}} & \cellcolor{red!10}{\textcolor{red}{.8445}} & \cellcolor{red!10}{\textcolor{red}{29.15}} & \cellcolor{red!10}{\textcolor{red}{.8063}} & \cellcolor{red!10}{\textcolor{red}{28.27}} & \cellcolor{red!10}{\textcolor{red}{.8538}} & \cellcolor{red!10}{\textcolor{red}{33.92}} & \cellcolor{red!10}{\textcolor{red}{.9460}}\\
    \midrule
    
    \multirow{10}*{\rotatebox{90}{$\times 4$}}
    & Bicubic & - & - & 28.42 & .8104 & 26.00 & .7027 & 25.96 & .6675 & 23.14 & .6577 & 24.89 & .7866\\
    & CARN-M~\cite{ahn2018carn} & 415K & 33 & 31.92 & .8903 & 28.42 & .7762 & 27.44 & .7304 & 25.62 & .7694 & - & - \\
    & IMDN~\cite{hui2019imdn} & 715K &41&32.21&.8948&28.58&.7811&27.56&.7353&26.04&.7838&30.46&.9075\\
    & PAN~\cite{zhao2020pan} & 272K & 28 & 32.13 & .8948 & 28.61 & .7822 & 27.59 & .7363 & 26.11 & .7854 & 30.51 & .9095\\
    & DRSAN~\cite{park2021drsan} & 410K & 31 & 32.15 & .8935 & 28.54 & .7813 & 27.54 & .7364 & 26.06 & .7858 & - & - \\
    & DDistill-SR~\cite{wang2022ddistill} & 434K & 33 & \cellcolor{blue!10}{\textcolor{blue}{32.23}} & \cellcolor{blue!10}{\textcolor{blue}{.8960}} & 28.62 & .7823 & 27.58 & .7365 & \cellcolor{blue!10}{\textcolor{blue}{26.20}} & \cellcolor{red!10}{\textcolor{red}{.7891}} & 30.48 & .9090 \\
    & ShuffleMixer~\cite{sun2022shufflemixer} & 411K & 28 & 32.21 & .8953 & \cellcolor{blue!10}{\textcolor{blue}{28.66}} & \cellcolor{blue!10}{\textcolor{blue}{.7827}} & \cellcolor{blue!10}{\textcolor{blue}{27.61}} & \cellcolor{blue!10}{\textcolor{blue}{.7366}} & 26.08 & .7835 & \cellcolor{blue!10}{\textcolor{blue}{30.65}} & \cellcolor{blue!10}{\textcolor{blue}{.9093}} \\
    & SAFMN~\cite{sun2023safmn} & 240K & 14 & 32.18 & .8948 & 28.60 & .7813 & 27.58 & .7359 & 25.97 & .7809 & 30.43 & .9063 \\
    & SeemoRe-T (\textit{ours}) & 232K &12 & \cellcolor{red!10}{\textcolor{red}{32.31}} & \cellcolor{red!10}{\textcolor{red}{.8965}} & \cellcolor{red!10}{\textcolor{red}{28.72}} & \cellcolor{red!10}{\textcolor{red}{.7840}} & \cellcolor{red!10}{\textcolor{red}{27.65}} & \cellcolor{red!10}{\textcolor{red}{.7384}} & \cellcolor{red!10}{\textcolor{red}{26.23}} & \cellcolor{blue!10}{\textcolor{blue}{.7883}} & \cellcolor{red!10}{\textcolor{red}{30.82}} & \cellcolor{red!10}{\textcolor{red}{.9107}}\\
    \bottomrule
    \end{tabularx}
\end{table*}

\begin{table*}[ht]
    \centering
    \footnotesize
    \fboxsep0.85pt
    \setlength\tabcolsep{2pt}   
    \caption{\textit{Comparison to lightweight SR Transformers.}~PSNR (dB $\uparrow$) and SSIM ($\uparrow$) metrics are reported on the Y-channel. \colorbox{red!10}{\textcolor{red}{Best}} and \colorbox{blue!10}{\textcolor{blue}{second best}} performances are highlighted. GMACS (G) are computed by upscaling to a $1280\times720$ HR image.
    SeemoRe-L outperforms or achieves comparable performance to compared Transformers while being more efficient. 
    $\times 3$ results are in the Supplemental.
    }
    \label{tab:exp:lightweight_sota}
    \begin{tabularx}{\textwidth}{lX*{14}{c}}
    \toprule
    & \multirow{2}{*}{Method} & \multirow{2}{*}{Params} & \multirow{2}{*}{GMACS} &  
    \multicolumn{2}{c}{Set5} & \multicolumn{2}{c}{Set14} & \multicolumn{2}{c}{BSD100} & \multicolumn{2}{c}{Urban100} & \multicolumn{2}{c}{Manga109} \\
    \cmidrule(lr){5-6} \cmidrule(lr){7-8} \cmidrule(lr){9-10} \cmidrule(lr){11-12} \cmidrule(lr){13-14} 
    & & & &PSNR&SSIM &PSNR&SSIM &PSNR&SSIM &PSNR&SSIM &PSNR&SSIM\\
    \midrule
    
    \multirow{8}*{\rotatebox{90}{$\times 2$}} 
    & Bicubic & - & - & 33.66 & .9299 & 30.24 & .8688 & 29.56 & .8431 & 26.88 & .8403 & 30.80 & .9339\\
    & SwinIR-Light~\cite{liang2021swinir} & 910K &244&38.14&.9611&33.86&.9206&32.31&.9012&32.76&.9340&39.12&.9783\\
    & ELAN-Light~\cite{zhang2022elan} & 621K &201 &38.17&.9611& \cellcolor{blue!10}{\textcolor{blue}{33.94}}&.9207&32.30&.9012&32.76&.9340&39.11&.9782\\
    & SRFormer-Light~\cite{zhou2023srformer} & 853K & 236 & 38.23 & .9613 &  \cellcolor{blue!10}{\textcolor{blue}{33.94}} & .9209 & \cellcolor{red!10}{\textcolor{red}{32.36}} & \cellcolor{red!10}{\textcolor{red}{.9019}} & \cellcolor{red!10}{\textcolor{red}{32.91}} & \cellcolor{red!10}{\textcolor{red}{.9353}} & \cellcolor{blue!10}{\textcolor{blue}{39.28}} & .9785 \\
    & ESRT~\cite{lu2022esrt} & 751K & 191 & 38.03&.9600&33.75&.9184&32.25&.9001&32.58&.9318&39.12&.9774 \\
    & SwinIR-NG~\cite{choi2023n} & 1181K & 274 & 38.17 & .9612 & 33.94 & .9205 & 32.31 & .9013 & 32.78 & .9340 & 39.20 & .9781\\
    & DAT-Light~\cite{chen2023dat} & 553K & 194 &  \cellcolor{blue!10}{\textcolor{blue}{38.24}} &  \cellcolor{blue!10}{\textcolor{blue}{.9614}} &  \cellcolor{red!10}{\textcolor{red}{34.01}} &  \cellcolor{red!10}{\textcolor{red}{.9214}} & 32.34 & \cellcolor{blue!10}{\textcolor{blue}{.9019}} & \cellcolor{blue!10}{\textcolor{blue}{32.89}} & \cellcolor{blue!10}{\textcolor{blue}{.9346}} & \cellcolor{red!10}{\textcolor{red}{39.49}} & \cellcolor{blue!10}{\textcolor{blue}{.9788}} \\
    & SeemoRe-L (\textit{ours}) & 931K& 197& \cellcolor{red!10}{\textcolor{red}{38.27}}	&\cellcolor{red!10}{\textcolor{red}{.9616}}	& \cellcolor{red!10}{\textcolor{red}{34.01}} & \cellcolor{blue!10}{\textcolor{blue}{.9210}}	& \cellcolor{blue!10}{\textcolor{blue}{32.35}}	&.9018 &32.87 &.9344	& \cellcolor{red!10}{\textcolor{red}{39.49}}	& \cellcolor{red!10}{\textcolor{red}{.9790}} \\

    \midrule

    \multirow{8}*{\rotatebox{90}{$\times 4$}} 
    & Bicubic & - & - & 28.42 & .8104 & 26.00 & .7027 & 25.96 & .6675 & 23.14 & .6577 & 24.89 & .7866\\
    & SwinIR-Light~\cite{liang2021swinir} & 897K& 64 &32.44&.8976&28.77&.7858&27.69&.7406&26.47&.7980&30.91&.9151 \\
    & ELAN-Light~\cite{zhang2022elan} & 601K & 54 &32.43&.8975&28.78&.7858&27.69&.7406&26.54&.7982&30.92&.9150\\
    & SRFormer-Light~\cite{zhou2023srformer} & 873K &63&\cellcolor{blue!10}{\textcolor{blue}{32.51}}&.8988&28.82 &.7872& 27.73 &\cellcolor{blue!10}{\textcolor{blue}{.7422}}& \cellcolor{blue!10}{\textcolor{blue}{26.67}}&.8032&31.17&.9165\\
    & ESRT~\cite{lu2022esrt} & 751K & 68 & 32.19&.8947&28.69&.7833&27.69&.7379&26.39&.7962&30.75&.9100\\
    & SwinIR-NG~\cite{choi2023n} & 1201K & 63 & 32.44 & .8980 & 28.83 & .7870 & 27.73 & .7418 & 26.61 & .8010 & 31.09 & .9161\\
    & DAT-Light~\cite{chen2023dat} & 573K & 50 &  \cellcolor{red!10}{32.57} &  \cellcolor{red!10}{\textcolor{red}{.8991}} &  \cellcolor{blue!10}{\textcolor{blue}{28.87}} & \cellcolor{blue!10}{\textcolor{blue}{.7879}} & \cellcolor{blue!10}{\textcolor{blue}{27.74}} & \cellcolor{red!10}{\textcolor{red}{.7428}} & 26.64 & \cellcolor{blue!10}{\textcolor{blue}{.8033}} & \cellcolor{blue!10}{\textcolor{blue}{31.37}} & \cellcolor{blue!10}{\textcolor{blue}{.9178}} \\
    & SeemoRe-L (\textit{ours}) & 969K & 50 &  \cellcolor{blue!10}{\textcolor{blue}{32.51}} &  \cellcolor{blue!10}{\textcolor{blue}{.8990}} &  \cellcolor{red!10}{\textcolor{red}{28.92}} &  \cellcolor{red!10}{\textcolor{red}{.7888}} &  \cellcolor{red!10}{\textcolor{red}{27.78}} &  \cellcolor{red!10}{\textcolor{red}{.7428}} &  \cellcolor{red!10}{\textcolor{red}{26.79}} &  \cellcolor{red!10}{\textcolor{red}{.8046}} &  \cellcolor{red!10}{\textcolor{red}{31.48}} &  \cellcolor{red!10}{\textcolor{red}{.9181}}\\
    \bottomrule
    \end{tabularx}
\end{table*}

\begin{figure*}[t]
    \centering
    \tiny
    \setlength\tabcolsep{0.75pt}
        \begin{tabularx}{\textwidth}{*{6}{C}}
            \multicolumn{2}{c}{\includegraphics[width=0.33\textwidth, trim=0 0 0 7.25cm, clip]{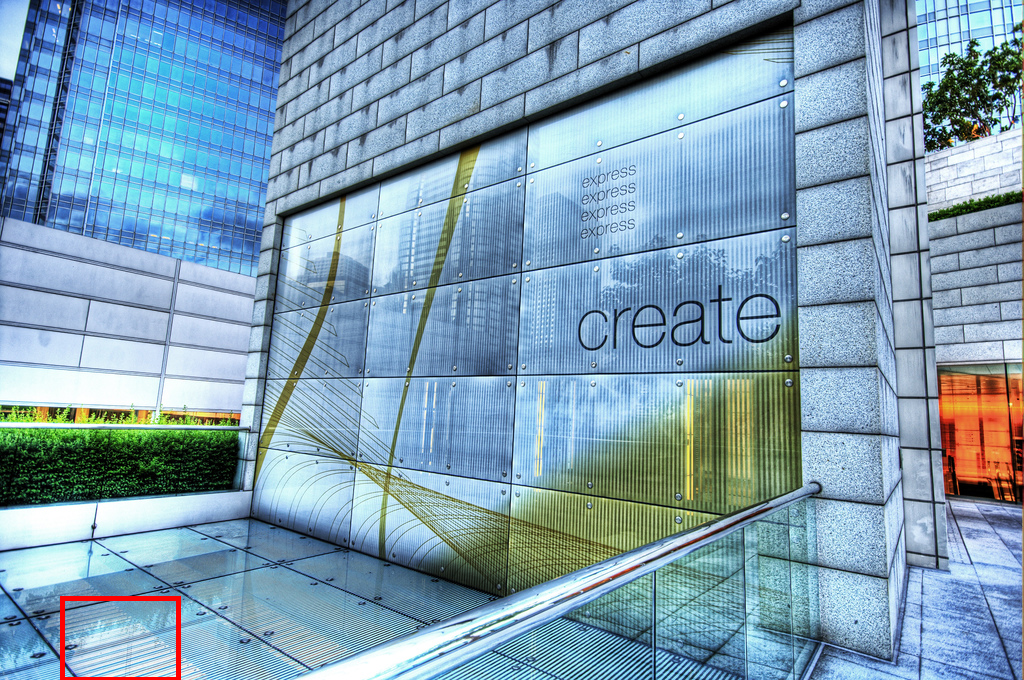}} &
            \multicolumn{2}{c}{\includegraphics[width=0.33\textwidth, trim=0 10.25cm 0 0, clip]{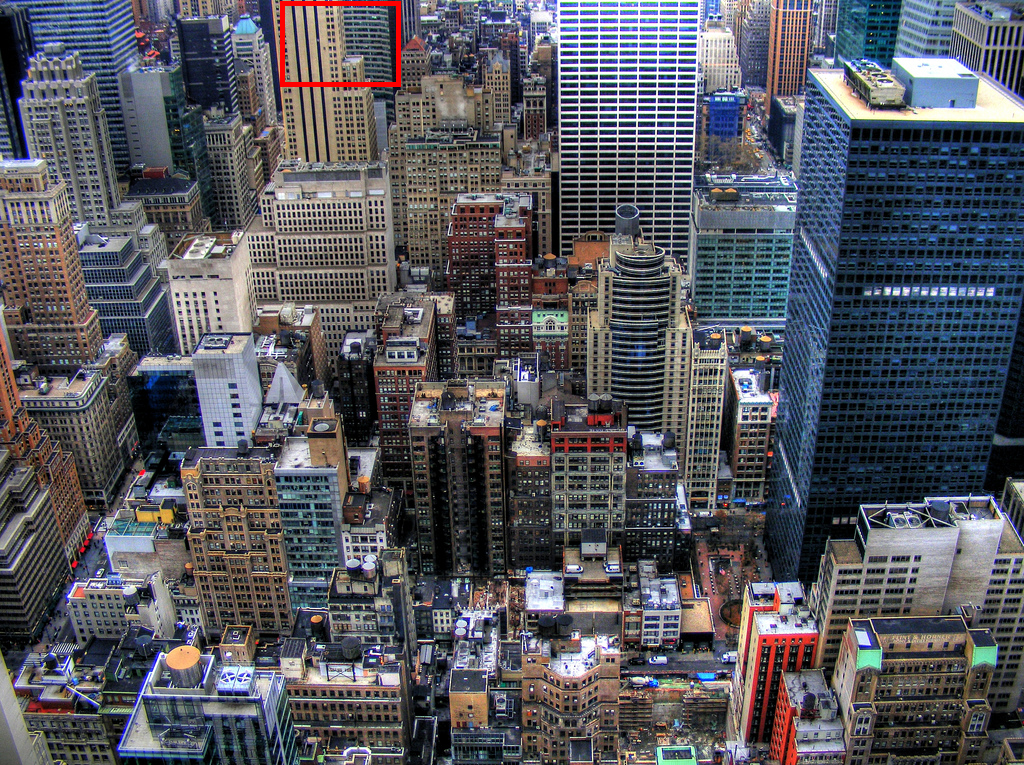}} &
            \multicolumn{2}{c}{\includegraphics[width=0.33\textwidth, trim=0 0 0 27cm, clip]{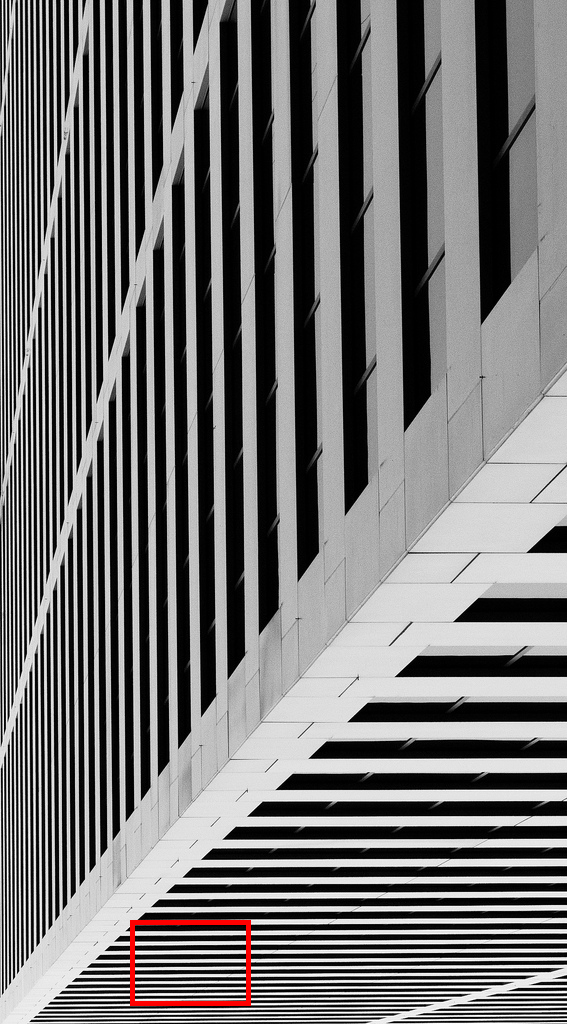}}  \\

            \multicolumn{2}{c}{Urban100: \text{img60} ($\times 4$)} &
            \multicolumn{2}{c}{Urban100: \text{img73} ($\times 4$)} &
            \multicolumn{2}{c}{Urban100: \text{img11} ($\times 4$)} \\

            \includegraphics[width=0.16\textwidth]{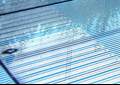} &
            \includegraphics[width=0.16\textwidth]{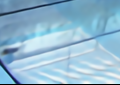} &
            \includegraphics[width=0.16\textwidth]{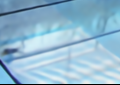} &
            \includegraphics[width=0.16\textwidth]{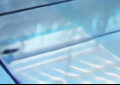} &
            \includegraphics[width=0.16\textwidth]{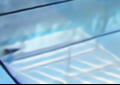} &
            \includegraphics[width=0.16\textwidth]{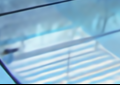} \\

            \includegraphics[width=0.16\textwidth]{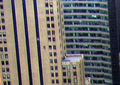} &
            \includegraphics[width=0.16\textwidth]{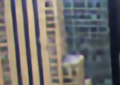} &
            \includegraphics[width=0.16\textwidth]{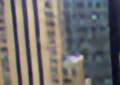} &
            \includegraphics[width=0.16\textwidth]{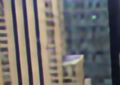} &
            \includegraphics[width=0.16\textwidth]{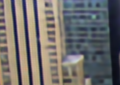} &
            \includegraphics[width=0.16\textwidth]{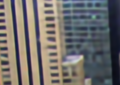} \\

            \includegraphics[width=0.16\textwidth]{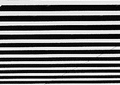} &
            \includegraphics[width=0.16\textwidth]{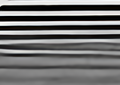} &
            \includegraphics[width=0.16\textwidth]{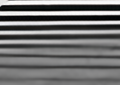} &
            \includegraphics[width=0.16\textwidth]{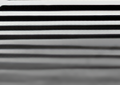} &
            \includegraphics[width=0.16\textwidth]{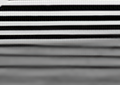} &
            \includegraphics[width=0.16\textwidth]{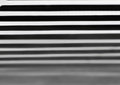} \\
            HR Crop & DDistill-SR & ShuffleMixer & SwinIR-Light &  DAT-Light & SeemoRe-L \\       
        \end{tabularx}
    \caption{Visual comparison of SeemoRe with state-of-the-art methods on challenging cases for $\times 4$ SR from the Urban100 benchmark.}
    \label{fig:exp:visuals}
\end{figure*}

\begin{table}[t]
    \centering
    \footnotesize
    \setlength\tabcolsep{2pt}
    \caption{\textit{Complexity Analysis.} Runtime (ms, $\downarrow$) and memory consumption (M, $\downarrow$) averaged across 200 samples using a NVIDIA RTX 4090 device.}
    \label{tab:exp:runtime}
    \begin{subtable}[l]{\columnwidth}
        \centering
        \setlength\tabcolsep{2pt}
         \begin{tabularx}{\columnwidth}{X*{4}{c}}
            \toprule
            Method & Input & Scale & Time & GPU Memory\\
            \midrule
            DAT-Light & \multirow{7}{*}{$[320, 180]$} & \multirow{7}{*}{$\times 4$} & 210.12 & 8715.1 \\
            SwinIR-Light & & & 131.25 & 6175.3\\
            SRFormer-Light & & & 103.95 & 7270.1\\
            SeemoRe-L (\textit{ours})& & & \textbf{17.99} & 9531.6 \\
            \midrule
            ShuffleMixer &  &  & 7.40 & \textbf{1380.7}\\
            DDistill-SR & & & 11.20 & 2822.1\\
            SeemoRe-T (\textit{ours})& &  &\textbf{5.66} & 2744.9 \\

            \bottomrule
        \end{tabularx} 
        \subcaption{Runtime and memory consumption.}
    \end{subtable}
    \begin{subtable}[l]{\columnwidth}
        \centering
        \setlength\tabcolsep{2pt}
        \label{tab:exp:tiny_transformers}
            \begin{tabularx}{\columnwidth}{lX*{4}{c}}
            \toprule
            Scale & Method & Params. & GMACS & Urban100 & Manga109 \\
            \midrule
            \multirow{4}{*}{\rotatebox{90}{$\times 2$}} & SwinIR\textsuperscript{$\ast$} &   191K & 48 & 31.56 & 38.07 \\
            & SRFormer\textsuperscript{$\ast$} &  188K & 49 & 31.60 & 38.59 \\
            & DAT$^\ast$ & 115K & 44 & 31.91 & 38.80 \\
            & \cellcolor{red!10}{SeemoRe-T} & \cellcolor{red!10}{\textbf{220K}} & \cellcolor{red!10}{\textbf{45}} & \cellcolor{red!10}{\textbf{32.22}} & \cellcolor{red!10}{\textbf{39.01}} \\
            \bottomrule
            \end{tabularx}
            \subcaption{PSNR (dB $\uparrow$) on the Y-Channel. $\ast$ denotes retrained models.}
    \end{subtable}
    \vspace{-4mm}
\end{table}

\paragraph{Datasets and Evaluation.} Following the SR literature~\cite{liang2021swinir,chen2023dat}, we utilize DIV2K~\cite{agustsson2017ntire} and Flickr2K~\cite{lim2017enhanced} datasets for training. We produce LR images using bicubic downscaling of HR images. When testing our method, we assess its performance on canonical benchmark datasets for SR - Set5~\cite{bevilacqua2012low}, Set14~\cite{zeyde2010single}, BSD100~\cite{martin2001database}, Urban100~\cite{huang2015single} and Manga109~\cite{matsui2017manga109}. We calculate PSNR and SSIM results on the Y-channel from the YCbCr color space.

\paragraph{Implementation Details.}
We augment our training data with randomly extracted $64\times64$-sized crops, with random rotation, horizontal and vertical flipping.  Similar to~\cite{sun2022shufflemixer,sun2023safmn}, we minimize the L1-Norm between SR output and HR ground truth in the pixel and frequency domain using Adam~\cite{kingma2017adam} optimizer for $500$K iterations with a batch size of $32$ and initial learning rate of $1\times10^{-3}$ halving it at following milestones: [$250K$,$400K$,$450K$,$475K$].  All experiments are conducted with the PyTorch framework on NVIDIA RTX 4090 GPUs.
We design our smallest model (SeemoRe-T) with 6 RGs. The feature dimension and channel expansion factor in GatedFFN are set to $36$ and $2$, respectively. For all MoRE sub-modules, we select an exponential growth of the channel dimensionality and choose in total of $3$ experts. The kernel size in SEE is set to $11 \times 11$. 
More details can be found in the supplemental, \textit{c.f.} \cref{tab:supp:arch_config}.

\subsection{Comparison to State-of-the-Art Methods}
We present quantitative results for $\times 2$, $\times 3$, and $\times 4$ image SR, comparing against current efficient state-of-the-art models in \cref{tab:exp:efficient_sota}, including CARN-M~\cite{ahn2018carn}, IMDN~\cite{hui2019imdn}, PAN~\cite{zhao2020pan}, DRSAN~\cite{park2021drsan}, DDistill-SR~\cite{wang2022ddistill}, ShuffleMixer~\cite{sun2022shufflemixer}, and SAFMN~\cite{sun2023safmn}. Additionally, we evaluate against lightweight variants of popular Transformer-based SR models such as SwinIR~\cite{liu2021swin}, ELAN~\cite{zhang2022elan}, and SRFormer~\cite{zhou2023srformer} in \cref{tab:exp:lightweight_sota}. 
Our proposed SeemoRe-T stands out as the most efficient method, consistently surpassing all other methods across all benchmarks and scale factors. For instance as clear from \cref{tab:exp:efficient_sota}, on the Urban100 and Manga109 benchmarks ($\times 2$), SeemoRe-T outperforms SAFMN~\cite{sun2023safmn} by $0.41$dB and $0.30$dB, respectively. Furthermore, with $47\%$ fewer parameters and $65\%$ fewer GMACS than DDistill-SR~\cite{wang2022ddistill}, SeemoRe-T achieves on average $0.12$ dB higher PSNR results across all benchmarks ($\times 4$). Scaling our method up to a comparable size with lightweight Transformers yields comparable or superior results. As demonstrated in \cref{tab:exp:lightweight_sota}, our SeemoRe-L outperforms SwinIR-Light and SRFormer-Light on Manga109 ($\times 4$) by $0.57$dB and $0.31$dB, while requiring fewer GMACS.

\paragraph{Visual Results.} We show visual comparisons ($\times 4$) in \cref{fig:exp:visuals}. In some challenging scenarios, the previous methods may suffer blurring artifacts, distortions, or inaccurate texture restoration. Contrary to others, our SeemoRe alleviates the blurring artifacts better and maintains structural fidelity. For instance, in image \textit{img60} and \textit{img73} from Urban100, certain methods like DDistill-SR, SwinIR-Light and DAT-Light fail to accurately reconstruct shadow patterns or window struts, whereas our method exhibits strong recovery of fine details.  These visual comparisons highlight SeemoRe's ability to reconstruct high-quality images by effectively leveraging local and contextual information. Coupled with quantitative comparisons, these findings underscore the effectiveness of our method. More visual results can be found in the Supplementary material. 

\subsection{Model Complexity Trade-Off}
In the vision domain, scalability becomes more paramount. We strive to expand the limits of our SeemoRe framework, optimizing for both reconstruction fidelity and efficiency. The framework provides three complexity scales—tiny (T), base (B), and large (L)—with progressively improved reconstruction performance, \textit{c.f} \cref{fig:exp:efficiency}.
In  \cref{tab:exp:runtime}, we present comparisons of memory usage and running time, demonstrating that our SeemoRe-T outperforms representative state-of-the-art methods. By using the low-rank feature modulation and simultaneous aggregation of the channel-spatial dependencies, the GPU consumption of our SeemoRe-T is $3\%$ less than DDistill-SR, while being $2$ times faster. Additionally, \cref{tab:exp:runtime} highlights the significant efficiency advantage of SeemoRe over lightweight Transformers. Further results are provided in the Supplemental.
To further underscore our method's capability, we align SwinIR-Light and SRFormer-Light with a size and computational demand similar to ours, followed by retraining these downsized networks using our schedule. The results presented in \cref{tab:exp:tiny_transformers} highlight that SeemoRe-T significantly outperforms both Transformer-based models by a considerable margin.

\begin{table}[t]
    \centering
    \footnotesize
    \caption{\textit{Ablation on Blocks.} GMACS ($\downarrow$) are computed by upscaling to a $1280\times720$ HR image. We show results for $\times 2$ upscaling.}
    \label{tap:exp:block_ablations}
    \begin{subtable}[l]{\columnwidth}
    \setlength\tabcolsep{1pt}
    \begin{tabularx}{\columnwidth}{X*{6}{c}}
        \toprule
        Method & RME & SME & Params. & GMACS & Urban100 & Manga109 \\
        \midrule
        Baseline & - & - & 157K & 35 & 31.61 & 38.55 \\
        \multirow{3}{*}{SeemoRe-T}& $\checkmark$ & - & 199K & 40 & 31.96 &	38.75 \\
         & -  & $\checkmark$ & 178K & 40 & 31.97 &	38.87 \\
         & \cellcolor{red!10}{$\checkmark$} & \cellcolor{red!10}{$\checkmark$} & \cellcolor{red!10}{\textbf{220K}} & \cellcolor{red!10}{\textbf{45}} & \cellcolor{red!10}{\textbf{32.22}} & \cellcolor{red!10}{\textbf{39.01}} \\
        \bottomrule
    \end{tabularx} 
    \subcaption{Contribution of components.}
    \label{tab:exp:arch_components}
    \end{subtable}
\hfill
\vspace{-1mm}
    \begin{subtable}[l]{\columnwidth}
        \setlength\tabcolsep{1.5pt}
        \begin{tabularx}{\columnwidth}{lX*{2}{c}}
        \toprule
        Method & Macro Expert Order & Urban100 & Manga109 \\
        \midrule
        \multirow{2}{*}{SeemoRe-T} & Spatial - Rank & 32.17 & 38.93 \\
        & \cellcolor{red!10}{Rank - Spatial} & \cellcolor{red!10}{\textbf{32.22}} & \cellcolor{red!10}{\textbf{39.01}} \\    
        \bottomrule
        \end{tabularx}
        \subcaption{Block order.}
        \label{tab:exp:block_ordering}
    \end{subtable}
\hfill
\vspace{-1mm}
    \begin{subtable}[l]{\columnwidth}
    \setlength\tabcolsep{1.5pt}
    \label{tab:exp:kernel}
    \begin{tabularx}{\columnwidth}{X*{5}{c}}
        \toprule
        Method & Kernel $k$ & Params. & GMACS & Urban100 & Manga109 \\
        \midrule
        \multirow{4 }{*}{SeemoRe-T}
        & 3 &  217K & 44 & 32.04 & 38.84 \\
        & 7 &  219K & 45 &  32.10 & 38.96 \\
        & \cellcolor{red!10}{11} &  \cellcolor{red!10}{\textbf{220K}} & \cellcolor{red!10}{\textbf{45}} & \cellcolor{red!10}{\textbf{32.22}} & \cellcolor{red!10}{\textbf{39.01}}  \\
        \bottomrule
    \end{tabularx}
    \subcaption{Kernel size ($k$) variation.}
    \end{subtable}
    \vspace{-4mm}
\end{table}

\begin{table}[t]
    \centering
    \footnotesize
    \setlength\tabcolsep{1pt}
    \caption{\textit{Ablation on MoRE.} Exponential growth yields best performance in terms of parameter counts and PSNR. $\#\mathcal{E}$ denotes the number of experts and \textit{Dim.} the rank dimensionality. We show results for $\times 2$ upscaling.}
        \begin{subtable}[l]{\columnwidth}
        \begin{tabularx}{\columnwidth}{lC*{5}{c}}
            \toprule
            \multicolumn{2}{c}{Method} &  $\#\mathcal{E}$  & Dim. & Params. &  Urban100 & Manga109 \\
            \midrule
            \multicolumn{2}{c}{SeemoRe-T} \\
            \cmidrule(lr){1-7}
            \multirow{3}{*}{\rotatebox{90}{Growth}}
                 & $i+2$ & 6 & 2,3,4,5,6,7 & 229K & 32.12  & 39.01\\
                 & $2*i+2$ & 4 & 2,4,6,8 & 224K & 32.11 & 38.98\\ 
                 & $2^{i}$ & 4 & 2,4,8,16 & 231K & 32.21 & \textbf{39.02} \\
            \cmidrule(lr){2-7}
             \multirow{3}{*}{\rotatebox{90}{$\#\mathcal{E}$}}& \multirow{3}{*}{$2^{i}$} &
                 \cellcolor{red!10}{$3$} & \cellcolor{red!10}{$2,4,8$} & \cellcolor{red!10}{\textbf{220K}} & \cellcolor{red!10}{\textbf{32.22}} & \cellcolor{red!10}{39.01} \\
                & & $2$ & $2,4$ &  214K & 32.19 & 39.00 \\
                & & $1$ & $2$ &  211K & 32.16 & 38.92 \\
            \bottomrule
        \end{tabularx}
        \subcaption{Low-rank expert design.}
        \label{tab:exp:more_ablation}
        \end{subtable}
        \hfill
        \begin{subtable}[l]{\columnwidth}
        \setlength\tabcolsep{2pt}
            \begin{tabularx}{\columnwidth}{lX*{2}{c}}
            \toprule
            Method & Recursive Steps $t$ & Urban100 & Manga109 \\
            \midrule
            \multirow{3}{*}{SeemoRe-T} & 1 & 32.10 & 38.99 \\
            & \cellcolor{red!10}{2} & \cellcolor{red!10}{\textbf{32.22}} & \cellcolor{red!10}{\textbf{39.01}} \\    
            & 3 & 31.04 & 38.13 \\
            \bottomrule
            \end{tabularx}
        \subcaption{Recursive step ($t$) variation.}
        \label{tab:exp:recursive_steps}
    \end{subtable}
\end{table}

\subsection{Ablation Study}
We conduct detailed studies on the components within our approach. All experiments are conducted on the $\times2$ setting.

\paragraph{Macro Architecture.}~As reported in \cref{tab:exp:arch_components}, we evaluate the effectiveness of our proposed key architectural components by comparing them with a baseline model consisting solely of depthwise and pointwise convolutions, more details in Supplemental. After adding the proposed modules into the baseline model, their is a notable and persistent improvement in the results. The incorporation of RME or SME results in improvements of $0.26$ dB or $0.32$ dB on Urban100 over the baseline, respectively. Although both modules individually outperform the baseline with only a marginal increase in parameters, alternating the insertion of both the modules within each RG fully unleashes the model's capabilities while enhancing the overall efficiency. Overall, our SeemoRe-T obtains a  compelling gain of $0.49$ dB and $0.38$ dB on Urban100 and Manga109, respectively. 
Moreover, \cref{tab:exp:block_ordering} empirically justifies the chosen block ordering, showcasing the Rank-Spatial macro order design's superiority over permuted Spatial-Rank macro order. This empirical evidence supplements the qualitative justifications in \cref{sec:discussion} regarding the individual importance of MoRE and SEE blocks.

\begin{figure*}[th]
    \centering
    \begin{subfigure}[c]{\columnwidth}
        \begin{tikzpicture}[baseline]
  \begin{axis}[
    width=0.62\columnwidth,
    height=0.4\columnwidth,
    tick label style={font=\scriptsize},
    title style={font=\scriptsize}, 
    label style={font=\scriptsize},
    xticklabels=\empty,
    xtick style={draw=none},
    xmajorticks=false,
    axis on top,  
    ybar,
    bar width=1pt,
    enlargelimits=0.15,
    xticklabels=\empty,
    xtick style={draw=none},
    title={$\times 2$ Urban100},
    title style={anchor=north, at={(0.5,1.075)},},
    ylabel={$\#$Images},
    ylabel style={anchor=center, at={(-0.15,0.47)},},
    symbolic x coords={1, 2, 3, 4, 5, 6},
    xtick=data,
    ymajorgrids=true,
    grid style=dashed,
    legend image post style={scale=0.5},
    legend style={at={(0.22,.75)}, draw=none, fill=none, anchor=south, legend columns=-1, font=\tiny,},
  ]
    \addplot[draw=cvprblue,fill=cvprblue] coordinates {(1,20) (2,7) (3,5) (4,4) (5,2) (6,100)};
    \addplot[draw=orange,fill=orange] coordinates {(1,62) (2,85) (3,0) (4,45) (5,5) (6,0)};
    \addplot[draw=magenta,fill=magenta] coordinates {(1,18) (2,8) (3,95) (4,51) (5,93) (6,0)};
    \legend{$\mathcal{E}_1$, $\mathcal{E}_2$, $\mathcal{E}_3$,}
  \end{axis}
\end{tikzpicture}
\hfill
\begin{tikzpicture}[baseline]
  \begin{axis}[
    width=0.62\columnwidth,
    height=0.4\columnwidth,
    tick label style={font=\scriptsize},
    title style={font=\scriptsize}, 
    label style={font=\scriptsize},
    xticklabels=\empty,
    xtick style={draw=none},
    xmajorticks=false,
    yticklabels=\empty,
    axis on top, 
    ybar,
    bar width=1pt,
    enlargelimits=0.15,
    legend image post style={scale=0.5},
    legend style={at={(0.25,.82)}, draw=none, fill=none, anchor=south, legend columns=-1, font=\tiny,},
    ylabel style={anchor=center, at={(-0.15,0.5)},},
    title={$\times 2$ Manga109},
    title style={anchor=north, at={(0.5,1.075)},},
    symbolic x coords={1, 2, 3, 4, 5, 6},
    xtick=data,
    ymajorgrids=true,
    grid style=dashed,
  ]
    \addplot[draw=cvprblue,fill=cvprblue] coordinates {(1,0) (2,7) (3,0) (4,0) (5,21) (6,106)};
    \addplot[draw=orange,fill=orange] coordinates {(1,101) (2,102) (3,0) (4,85) (5,78) (6,2)};
    \addplot[draw=magenta,fill=magenta] coordinates {(1,8) (2,0) (3,109) (4,24) (5,10) (6,1)};

  \end{axis}
\end{tikzpicture}

\begin{tikzpicture}[baseline]
  \begin{axis}[
    width=0.62\columnwidth,
    height=0.4\columnwidth,
    xtick pos=left,
    ytick pos=left,
    axis on top,  
    ybar,
    bar width=1pt,
    enlargelimits=0.15,
    xlabel={Residual Groups},
    title={$\times 4$ Urban100},
    title style={anchor=north, at={(0.5,1.075)},},
    ylabel={$\#$Images},
    ylabel style={anchor=center, at={(-0.15,0.47)},},
    symbolic x coords={1, 2, 3, 4, 5, 6},
    xtick=data,
    ymajorgrids=true,
    grid style=dashed,
    tick label style={font=\scriptsize},
    title style={font=\scriptsize}, 
    label style={font=\scriptsize},
    tick align=inside,
  ]
    \addplot[draw=cvprblue,fill=cvprblue] coordinates {(1,43) (2,1) (3,26) (4,13) (5,98) (6,100)};
    \addplot[draw=orange,fill=orange] coordinates {(1,46) (2,12) (3,70) (4,18) (5,0) (6,0)};
    \addplot[draw=magenta,fill=magenta] coordinates {(1,11) (2,87) (3,4) (4,69) (5,2) (6,0)};
  \end{axis}
\end{tikzpicture}
\hfill
\begin{tikzpicture}[baseline]
  \begin{axis}[
    width=0.62\columnwidth, 
    height=0.4\columnwidth,
    xtick pos=left,
    ytick pos=left,
    xlabel near ticks,
    ylabel near ticks,
    yticklabels=\empty,
    axis on top, 
    ybar,
    bar width=1pt,
    enlargelimits=0.15,
    legend image post style={scale=0.5},
    legend style={at={(0.2,.82)}, draw=none, fill=none, anchor=south, legend columns=-1, font=\tiny,},
    xlabel={Residual Groups},
    ylabel style={anchor=center, at={(-0.05,0.5)},},
    title={$\times 4$ Manga109},
    title style={anchor=north, at={(0.5,1.075)},},
    symbolic x coords={1, 2, 3, 4, 5, 6},
    xtick=data,
    ymajorgrids=true,
    grid style=dashed,
    tick label style={font=\scriptsize},
    title style={font=\scriptsize}, 
    label style={font=\scriptsize},
    tick align=inside,
  ]
    \addplot[draw=cvprblue,fill=cvprblue] coordinates {(1,21) (2,2) (3,0) (4,0) (5,108) (6,109)};
    \addplot[draw=orange,fill=orange] coordinates {(1,87) (2,69) (3,109) (4,50) (5,0) (6,0)};
    \addplot[draw=magenta,fill=magenta] coordinates {(1,1) (2,38) (3,0) (4,59) (5,1) (6,0)};

  \end{axis}
\end{tikzpicture}
        \subcaption{}
        \label{fig:exp:routing}
    \end{subfigure}
    \hfill
    \begin{subfigure}[c]{\columnwidth}
        \input{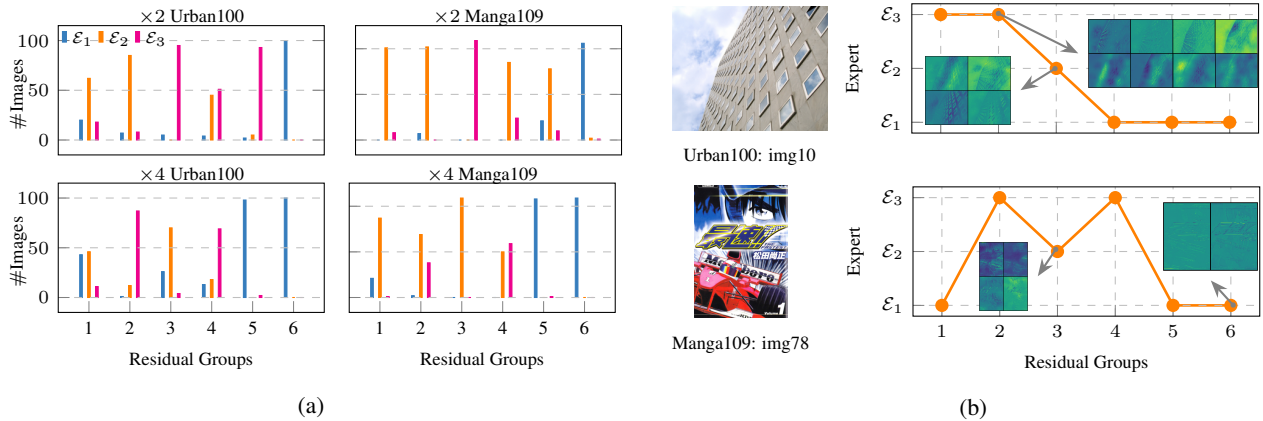}
        \subcaption{}
        \label{fig:exp:lowrank_vis}
    \end{subfigure}
    \vspace{-2mm}
    \caption{\textit{Low-Rank Analysis.} (a)~We plot the decisions made by the routing function for SeemoRe-T over the depth of the network.(b)~We visualize the low-rank features of SeemoRe-T for $\times 4$ SR given example images from Urban100 and Manga109.}
\end{figure*}

\begin{figure}[th]
    \centering
        \begin{tikzpicture}

        \node (image1) at (0,0) {\includegraphics[width=0.24\columnwidth, keepaspectratio]{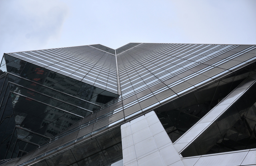}};
        \node (image2) at (0.242\columnwidth,0) {\includegraphics[width=0.24\columnwidth, keepaspectratio]{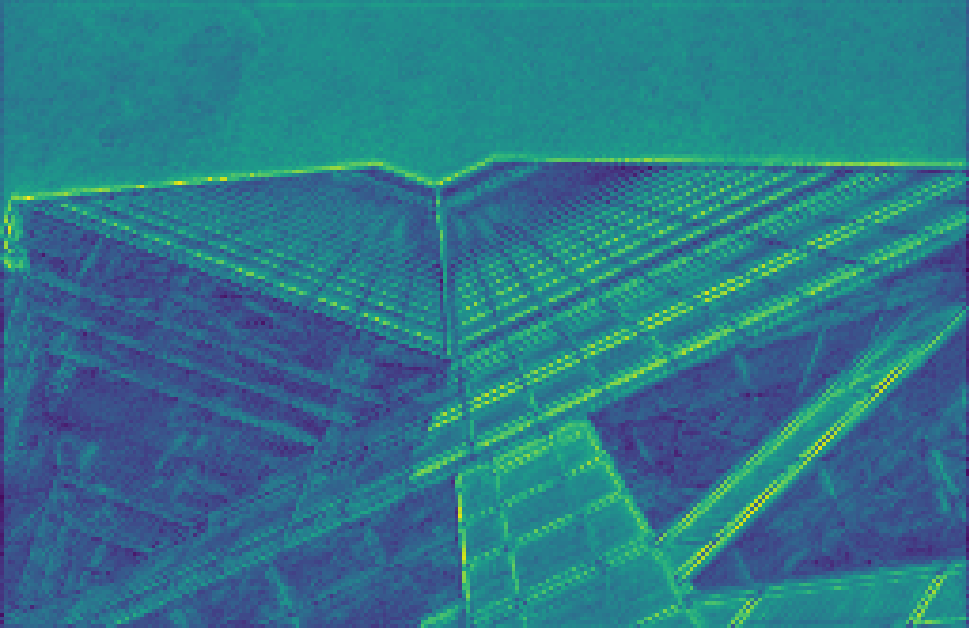}};
        \node (image3) at (0.484\columnwidth,0) {\includegraphics[width=0.24\columnwidth, keepaspectratio]{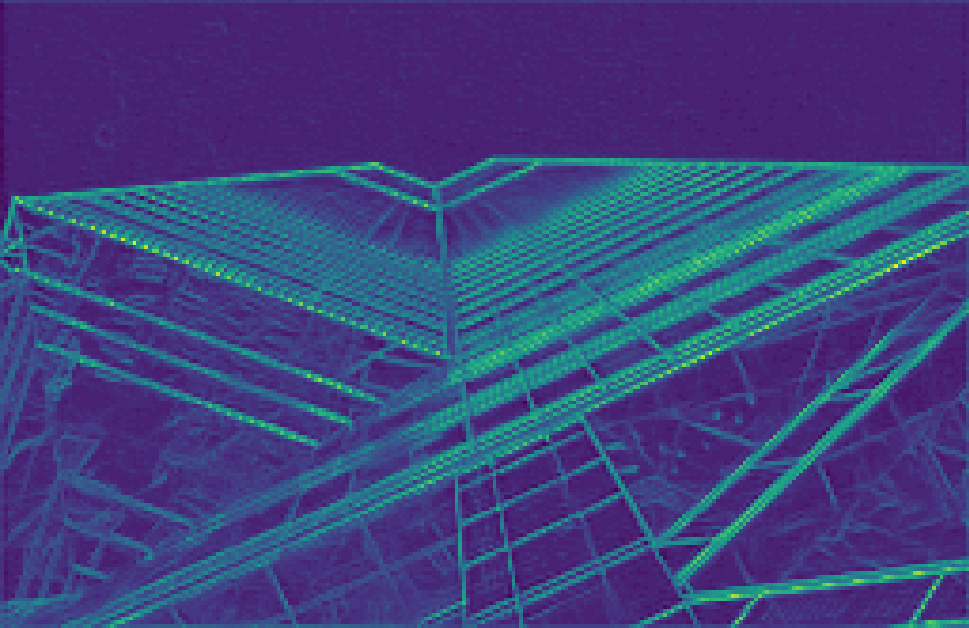}};
        \node (image4) at (0.726\columnwidth,0) {\includegraphics[width=0.24\columnwidth, keepaspectratio]{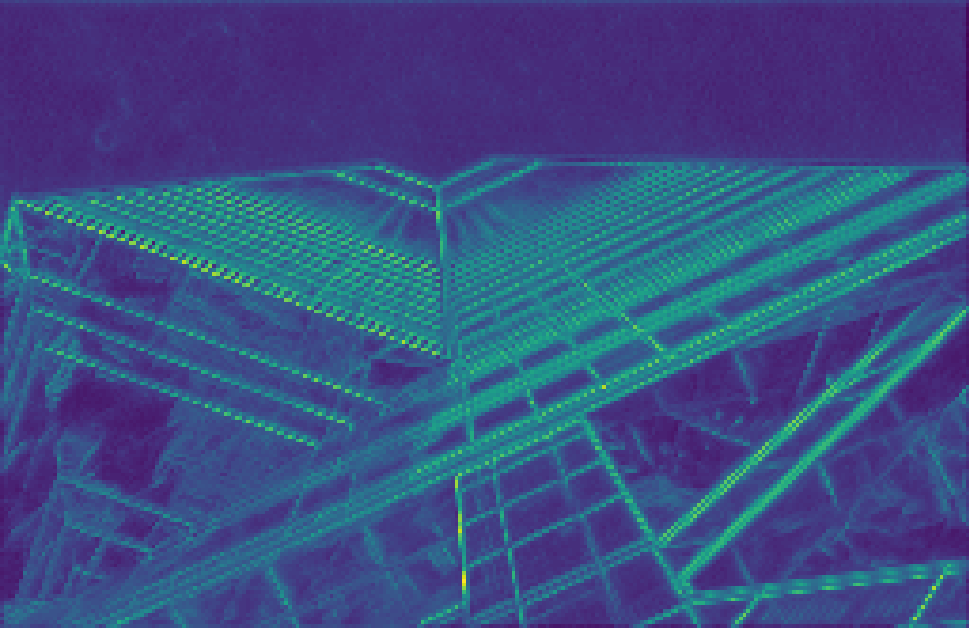}};

         \node (image5) at (0,-0.16\columnwidth) {\includegraphics[width=0.24\columnwidth, keepaspectratio]{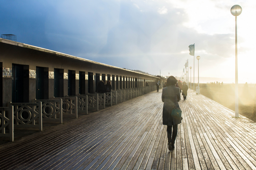}};
         
        \node (image6) at (0.242\columnwidth,-0.16\columnwidth) {\includegraphics[width=0.24\columnwidth, keepaspectratio]{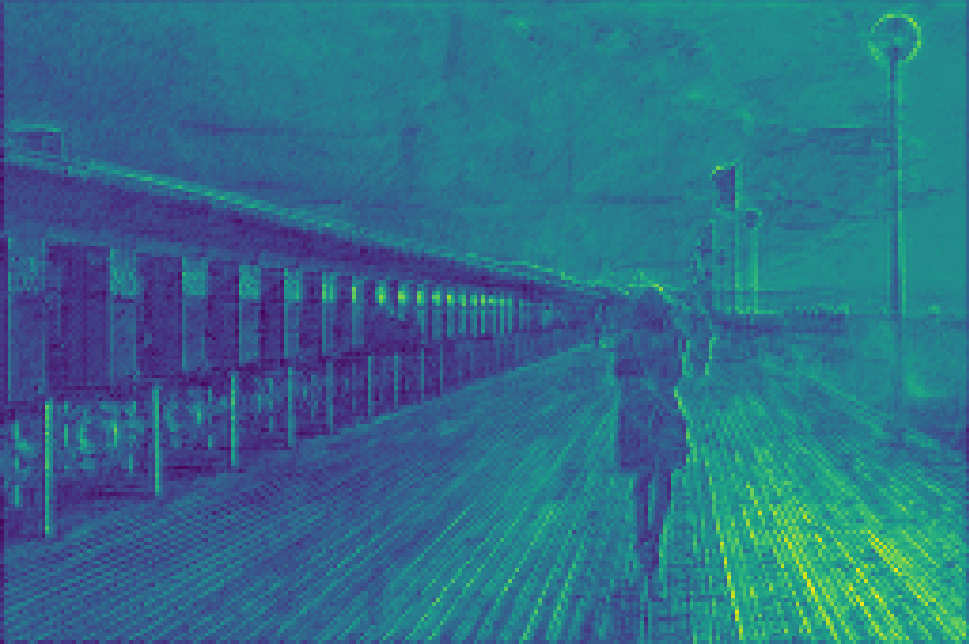}};
        \node (image7) at (0.484\columnwidth,-0.16\columnwidth) {\includegraphics[width=0.24\columnwidth, keepaspectratio]{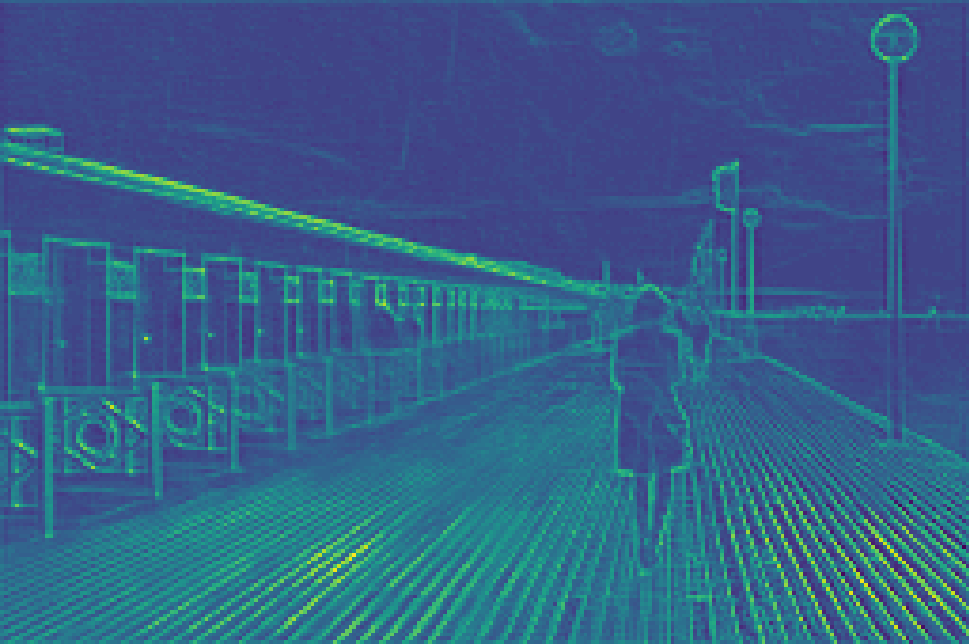}};
        \node (image8) at (0.726\columnwidth,-0.16\columnwidth) {\includegraphics[width=0.24\columnwidth, keepaspectratio]{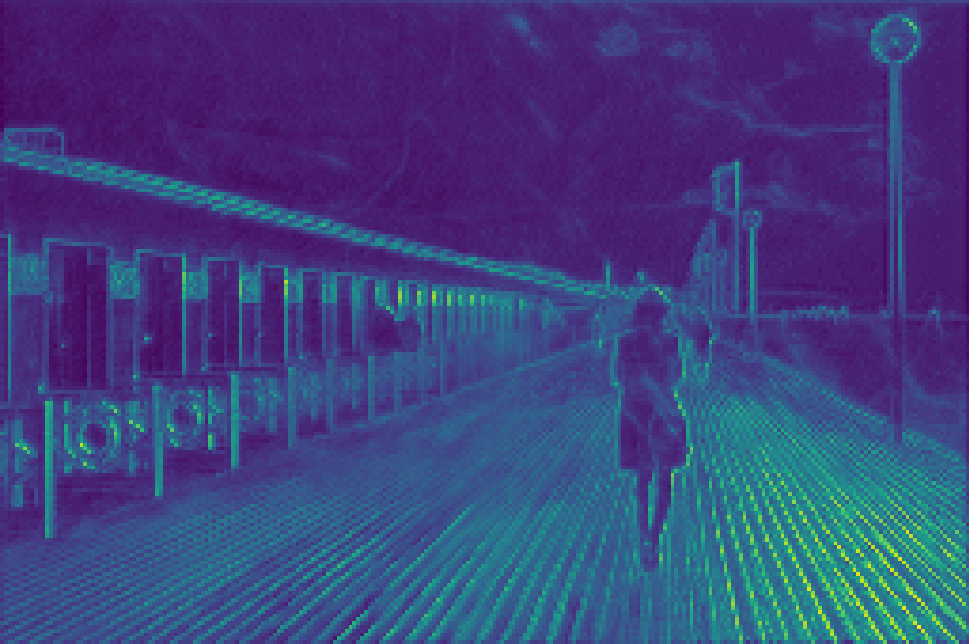}};

        \node[below] at (image5.south) {\scriptsize Input};
        \node[below] at (image6.south) {\scriptsize Before MoRE};
        \node[below] at (image7.south) {\scriptsize After MoRE};
        \node[below] at (image8.south) {\scriptsize After SEE};

        \draw[-stealth, red, thick] (3.5,0.5) -- (3.9,0.2);  
        \draw[-stealth, red, thick] (5.5,0.5) -- (5.9,0.2);

        \draw[-stealth, red, thick] (4.3,-1.9) -- (4.6,-1.6); 
        \draw[-stealth, red, thick] (6.3,-1.9) -- (6.6,-1.6); 
    \end{tikzpicture}
    \vspace{-6mm}
    \caption{\textit{Feature Visualization.} We present visualizations of feature maps before and after our proposed modules. Clearly, our MoRE block notably enhances activation sharpness via contextual feature modulation. Moreover, our SEE module improves learned representations by integrating spatial cues effectively.}
    \label{fig:exp:feature_vis}
    \vspace{-2mm}
\end{figure}

\paragraph{Design choices of SME.}~The main component of SME module, SEE deploys striped convolutions with large-kernel sizes to effectively module the spatial cues. \cref{tab:exp:kernel} demonstrates that deploying large kernel sizes improves the overall performance of the model. In particular, the PSNR shows a notable gain of 0.18 dB on Urban100 dataset when increasing the kernel size from 3$\times$3 to 11$\times$11 (keeping other settings intact), with only 3K increase in parameters.  It clearly proves that such a design benefits in the efficient use of the relevant information to augment the restoration of sharp regions spatially.

\paragraph{Design choices of RME.}~We motivate our design choices for the MoRE module in RME by varying the growth function and the number of experts as depicted in  \cref{tab:exp:more_ablation}.
When pursuing a dynamic solution for determining the optimal low-rank dimensionality, it becomes necessary to design the corresponding search space. First, we present results for $\times 2$ upscaling on Urban100 and Manga109 using different growth functions. Based on the observed outcomes, it is evident that an exponentially increasing low-rank dimensionality yields the best performance with marginal increase in the parameters. Hence, we opt to retain this search space design in all further experimentation. Next, we analyze the reconstruction quality based on the number of experts in each MoRE module, while exponentially increasing the low-rank dimensionality. Based on these experiments, we assert that the efficient results are obtained when we have three, as our total number of experts. 
Further, we ablate the choice of recursive steps for SeemoRe-T in \cref{tab:exp:recursive_steps}, where our plain version takes ($t = 2$). It can be seen that lower ($t = 1$) and higher ($t = 3$) values either fail to capture sufficient contextual information or overly compromise spatial image features.

\section{Discussion on Experts}
\label{sec:discussion}
Our model integrates experts at varying levels, each specializing in crucial factors for SR. In this section, we aim to elucidate their expertise.

\paragraph{Mixture of Low-Rank Experts.}~The decision-making process of the router at different network depths is illustrated in \cref{fig:exp:routing}. Notably, earlier blocks showcase a diverse range of rank choices ( $\mathcal{E}_1$,$\mathcal{E}_2$,$\mathcal{E}_3$), while deeper layers tend to favor lower ranks ($\mathcal{E}_1$) (Please note that for every $\mathcal{E}_i$, the corresponding rank dimension is $2^i$). This phenomenon can be attributed to the hierarchical feature learning nature of deep neural networks, aligning with our expectations. In fact, earlier layers typically capture low-level details and, at times, unwanted noise while reconstructing details in the input LR image, thus resulting in wide variations in the rank choices. In contrast, deeper layers focus on the main structures and key features required for SR. Hence, higher ranks at deeper layers are less favored, as they may introduce redundancy or noise that does not significantly contribute to the overall quality of the reconstructed image.
This design aspect provides our method with the flexibility to adapt to the complexity of the task, a capability that, to the best of our knowledge, has not yet been explored in the image reconstruction community. In \cref{fig:exp:lowrank_vis}, we further visualize the routing decisions and the corresponding low-rank feature maps for two exemplary input images. It is noteworthy that each individual rank carries distinct information while being mutually complementary. As the model depth increases, the network becomes proficient in restructuring these representations.

\paragraph{How important are MoRE and SEE?}~To substantiate the significance of the proposed MoRE and SEE modules, we analyze the feature maps before and after integrating both blocks into the RME module as depicted in \cref{fig:exp:feature_vis}. This analysis vividly showcases the advantages of leveraging MoRE for contextual information mining within RME. Notably, the activations exhibit reduced noise and enhanced sharpness. Additionally, we observe a synergistic interaction between MoRE and SEE at marked locations (indicated by red arrows): MoRE effectively refines global textures by filtering out noise, while SEE supplements over-filtered regions with critical local details.
\section{Conclusion}

We propose a novel ConvNet, named \textbf{S}eemo\textbf{R}e, for efficient and accurate image super-resolution. Our \textbf{S}eemo\textbf{R}e excels in modeling local and contextual information, surpassing both previous CNN-based and lightweight Transformer approaches in terms of efficiency and reconstruction fidelity. Unlike other approaches, we empirically demonstrate both the scalability of efficiency and reconstruction performance. In our approach, we intricately design the rank modulation expert to discern the most pivotal features, enhancing this compressed representation with valuable contextual cues. Our spatial enhancement expert efficiently integrates local spatial-wise information, unlocking the full potential of our architecture. This novel approach optimally exploits the low information regime in the input image, enhancing detail reconstruction while improving efficiency. Extensive experiments on image super-resolution demonstrate that our proposed \textbf{S}eemo\textbf{R}e achieves consistent superior performance over recent state-of-the-art efficient methods on all considered SR benchmarks, while even being on par with the lightweight Transformers in terms of reconstruction fidelity.

\section*{Impact Statement}
This paper presents work whose goal is to advance the field of Machine Learning, specifically efficient image super-resolution. There are many potential societal consequences of our work, none which we feel must be specifically highlighted here. However, applying super-resolution methods in AI-assisted software raises ethical concerns about privacy invasion and increased surveillance capabilities. Adherence to transparency, accountability, and privacy rights is crucial to mitigate potential harm and ensure responsible deployment in alignment with societal values.

\section*{Acknowledgments}
This work was supported by The Alexander von Humboldt Foundation.

\bibliography{main}

\begin{thebibliography}{52}
\providecommand{\natexlab}[1]{#1}
\providecommand{\url}[1]{\texttt{#1}}
\expandafter\ifx\csname urlstyle\endcsname\relax
  \providecommand{\doi}[1]{doi: #1}\else
  \providecommand{\doi}{doi: \begingroup \urlstyle{rm}\Url}\fi

\bibitem[Agustsson \& Timofte(2017)Agustsson and Timofte]{agustsson2017ntire}
Agustsson, E. and Timofte, R.
\newblock {NTIRE} 2017 challenge on single image super-resolution: Dataset and study.
\newblock In \emph{Proceedings of the IEEE Conference on Computer Vision and Pattern Recognition Workshops}, pp.\  126--135, 2017.

\bibitem[Ahn et~al.(2018)Ahn, Kang, and Sohn]{ahn2018carn}
Ahn, N., Kang, B., and Sohn, K.-A.
\newblock Fast, accurate, and lightweight super-resolution with cascading residual network.
\newblock In \emph{Proceedings of the European conference on computer vision (ECCV)}, pp.\  252--268, 2018.

\bibitem[Bengio et~al.(2013)Bengio, L{\'e}onard, and Courville]{bengio2013estimating}
Bengio, Y., L{\'e}onard, N., and Courville, A.
\newblock Estimating or propagating gradients through stochastic neurons for conditional computation.
\newblock \emph{arXiv preprint arXiv:1308.3432}, 2013.

\bibitem[Bevilacqua et~al.(2012)Bevilacqua, Roumy, Guillemot, and Alberi-Morel]{bevilacqua2012low}
Bevilacqua, M., Roumy, A., Guillemot, C., and Alberi-Morel, M.~L.
\newblock Low-complexity single-image super-resolution based on nonnegative neighbor embedding.
\newblock In \emph{Proceedings of the British Machine Vision Conference}, 2012.

\bibitem[Chen et~al.(2023)Chen, Zhang, Gu, Kong, Yang, and Yu]{chen2023dat}
Chen, Z., Zhang, Y., Gu, J., Kong, L., Yang, X., and Yu, F.
\newblock Dual aggregation transformer for image super-resolution.
\newblock In \emph{Proceedings of the IEEE/CVF International Conference on Computer Vision}, 2023.

\bibitem[Choi et~al.(2023)Choi, Lee, and Yang]{choi2023n}
Choi, H., Lee, J., and Yang, J.
\newblock N-gram in swin transformers for efficient lightweight image super-resolution.
\newblock In \emph{Proceedings of the IEEE/CVF Conference on Computer Vision and Pattern Recognition}, pp.\  2071--2081, 2023.

\bibitem[Conde et~al.(2023)Conde, Zamfir, and Timofte]{conde2023rtsr}
Conde, M.~V., Zamfir, E., and Timofte, R.
\newblock Efficient deep models for real-time 4k image super-resolution. ntire 2023 benchmark and report.
\newblock In \emph{Proceedings of the IEEE/CVF Conference on Computer Vision and Pattern Recognition (CVPR) Workshops}, pp.\  1495--1521, June 2023.

\bibitem[Dong et~al.(2014)Dong, Loy, He, and Tang]{dong2014learning}
Dong, C., Loy, C.~C., He, K., and Tang, X.
\newblock Learning a deep convolutional network for image super-resolution.
\newblock In \emph{Proceeding of the European Conference on Computer Vision}, pp.\  184--199. Springer, 2014.

\bibitem[Dosovitskiy et~al.(2021)Dosovitskiy, Beyer, Kolesnikov, Weissenborn, Zhai, Unterthiner, Dehghani, Minderer, Heigold, Gelly, Uszkoreit, and Houlsby]{dosovitskiy2021an}
Dosovitskiy, A., Beyer, L., Kolesnikov, A., Weissenborn, D., Zhai, X., Unterthiner, T., Dehghani, M., Minderer, M., Heigold, G., Gelly, S., Uszkoreit, J., and Houlsby, N.
\newblock An image is worth 16x16 words: Transformers for image recognition at scale.
\newblock In \emph{International Conference on Learning Representations}, 2021.

\bibitem[Duchon(1979)]{LanczosFilteringinOneandTwoDimensions}
Duchon, C.~E.
\newblock Lanczos filtering in one and two dimensions.
\newblock \emph{Journal of Applied Meteorology and Climatology}, 1979.

\bibitem[Hou et~al.(2022)Hou, Lu, Cheng, and Feng]{hou2022conv2former}
Hou, Q., Lu, C.-Z., Cheng, M.-M., and Feng, J.
\newblock Conv2former: A simple transformer-style convnet for visual recognition.
\newblock \emph{Proceedings of the IEEE Conference on Computer Vision and Pattern Recognition}, 2022.

\bibitem[Huang et~al.(2015)Huang, Singh, and Ahuja]{huang2015single}
Huang, J.-B., Singh, A., and Ahuja, N.
\newblock Single image super-resolution from transformed self-exemplars.
\newblock In \emph{Proceedings of the IEEE Conference on Computer Vision and Pattern Recognition}, pp.\  5197--5206, 2015.

\bibitem[Hui et~al.(2019)Hui, Gao, Yang, and Wang]{hui2019imdn}
Hui, Z., Gao, X., Yang, Y., and Wang, X.
\newblock Lightweight image super-resolution with information multi-distillation network.
\newblock In \emph{Proceedings of the ACM International Conference on Multimedia}, pp.\  2024--2032, 2019.

\bibitem[Ignatov et~al.(2021)Ignatov, Timofte, Denna, and Younes]{ignatov2021real}
Ignatov, A., Timofte, R., Denna, M., and Younes, A.
\newblock Real-time quantized image super-resolution on mobile npus, mobile ai 2021 challenge: Report.
\newblock In \emph{Proceedings of the IEEE/CVF Conference on Computer Vision and Pattern Recognition}, pp.\  2525--2534, 2021.

\bibitem[Ignatov et~al.(2023)Ignatov, Timofte, Denna, Younes, Gankhuyag, Huh, Kim, Yoon, Moon, Lee, et~al.]{ignatov2023efficient}
Ignatov, A., Timofte, R., Denna, M., Younes, A., Gankhuyag, G., Huh, J., Kim, M.~K., Yoon, K., Moon, H.-C., Lee, S., et~al.
\newblock Efficient and accurate quantized image super-resolution on mobile npus, mobile ai \& aim 2022 challenge: report.
\newblock In \emph{Computer Vision--ECCV 2022 Workshops: Tel Aviv, Israel, October 23--27, 2022, Proceedings, Part III}, pp.\  92--129. Springer, 2023.

\bibitem[Khani et~al.(2021)Khani, Sivaraman, and Alizadeh]{khani2021efficient}
Khani, M., Sivaraman, V., and Alizadeh, M.
\newblock Efficient video compression via content-adaptive super-resolution.
\newblock In \emph{Proceedings of the IEEE/CVF International Conference on Computer Vision}, pp.\  4521--4530, 2021.

\bibitem[Kim et~al.(2016)Kim, Lee, and Lee]{kim2016deeply}
Kim, J., Lee, J.~K., and Lee, K.~M.
\newblock Deeply-recursive convolutional network for image super-resolution.
\newblock In \emph{Proceedings of the IEEE conference on computer vision and pattern recognition}, pp.\  1637--1645, 2016.

\bibitem[Kingma \& Ba(2017)Kingma and Ba]{kingma2017adam}
Kingma, D.~P. and Ba, J.
\newblock Adam: A method for stochastic optimization, 2017.

\bibitem[Kong et~al.(2022)Kong, Li, Liu, Liu, He, Bai, Chen, and Fu]{kong2022rlfn}
Kong, F., Li, M., Liu, S., Liu, D., He, J., Bai, Y., Chen, F., and Fu, L.
\newblock Residual local feature network for efficient super-resolution.
\newblock In \emph{Proceedings of the European Conference on Computer Vision Workshops}, 2022.

\bibitem[Li et~al.(2022)Li, Zhang, Timofte, Van~Gool, Kong, Li, Liu, Du, Liu, Zhou, et~al.]{li2022ntire}
Li, Y., Zhang, K., Timofte, R., Van~Gool, L., Kong, F., Li, M., Liu, S., Du, Z., Liu, D., Zhou, C., et~al.
\newblock Ntire 2022 challenge on efficient super-resolution: Methods and results.
\newblock In \emph{Proceedings of the IEEE/CVF Conference on Computer Vision and Pattern Recognition}, pp.\  1062--1102, 2022.

\bibitem[Li et~al.(2023)Li, Zhang, Van~Gool, Timofte, et~al.]{li2023ntire_esr}
Li, Y., Zhang, Y., Van~Gool, L., Timofte, R., et~al.
\newblock {NTIRE} 2023 challenge on efficient super-resolution: Methods and results.
\newblock In \emph{Proceedings of the IEEE/CVF Conference on Computer Vision and Pattern Recognition Workshops}, 2023.

\bibitem[Liang et~al.(2021)Liang, Cao, Sun, Zhang, Van~Gool, and Timofte]{liang2021swinir}
Liang, J., Cao, J., Sun, G., Zhang, K., Van~Gool, L., and Timofte, R.
\newblock Swinir: Image restoration using swin transformer.
\newblock In \emph{Proceedings of the IEEE/CVF International Conference on Computer Vision}, 2021.

\bibitem[Liang et~al.(2022)Liang, Zeng, and Zhang]{liang2022efficient}
Liang, J., Zeng, H., and Zhang, L.
\newblock Efficient and degradation-adaptive network for real-world image super-resolution.
\newblock In \emph{European Conference on Computer Vision}, pp.\  574--591. Springer, 2022.

\bibitem[Lim et~al.(2017)Lim, Son, Kim, Nah, and Lee]{lim2017enhanced}
Lim, B., Son, S., Kim, H., Nah, S., and Lee, K.~M.
\newblock Enhanced deep residual networks for single image super-resolution.
\newblock In \emph{Proceedings of the IEEE Conference on Computer Vision and Pattern Recognition Workshops}, 2017.

\bibitem[Liu et~al.(2020{\natexlab{a}})Liu, Tang, and Wu]{liu2020rfdn}
Liu, J., Tang, J., and Wu, G.
\newblock Residual feature distillation network for lightweight image super-resolution.
\newblock In \emph{Proceedings of the European Conference on Computer Vision Workshops}, 2020{\natexlab{a}}.

\bibitem[Liu et~al.(2020{\natexlab{b}})Liu, Zhang, Tang, Tang, and Wu]{liu2020residual}
Liu, J., Zhang, W., Tang, Y., Tang, J., and Wu, G.
\newblock Residual feature aggregation network for image super-resolution.
\newblock In \emph{Proceedings of the IEEE/CVF conference on computer vision and pattern recognition}, pp.\  2359--2368, 2020{\natexlab{b}}.

\bibitem[Liu et~al.(2021)Liu, Lin, Cao, Hu, Wei, Zhang, Lin, and Guo]{liu2021swin}
Liu, Z., Lin, Y., Cao, Y., Hu, H., Wei, Y., Zhang, Z., Lin, S., and Guo, B.
\newblock Swin transformer: Hierarchical vision transformer using shifted windows.
\newblock In \emph{Proceedings of the IEEE International Conference on Computer Vision}, 2021.

\bibitem[Lu et~al.(2022)Lu, Li, Liu, Huang, Zhang, and Zeng]{lu2022esrt}
Lu, Z., Li, J., Liu, H., Huang, C., Zhang, L., and Zeng, T.
\newblock Transformer for single image super-resolution.
\newblock In \emph{Proceedings of the IEEE/CVF Conference on Computer Vision and Pattern Recognition Workshops}, 2022.

\bibitem[Martin et~al.(2001)Martin, Fowlkes, Tal, and Malik]{martin2001database}
Martin, D., Fowlkes, C., Tal, D., and Malik, J.
\newblock A database of human segmented natural images and its application to evaluating segmentation algorithms and measuring ecological statistics.
\newblock In \emph{Proceedings of the IEEE International Conference on Computer Vision}, volume~2, pp.\  416--423. IEEE, 2001.

\bibitem[Matsui et~al.(2017)Matsui, Ito, Aramaki, Fujimoto, Ogawa, Yamasaki, and Aizawa]{matsui2017manga109}
Matsui, Y., Ito, K., Aramaki, Y., Fujimoto, A., Ogawa, T., Yamasaki, T., and Aizawa, K.
\newblock Sketch-based manga retrieval using manga109 dataset.
\newblock \emph{Multimedia Tools and Applications}, 2017.

\bibitem[Park et~al.(2021)Park, Soh, and Cho]{park2021drsan}
Park, K., Soh, J.~W., and Cho, N.~I.
\newblock Dynamic residual self-attention network for lightweight single image super-resolution.
\newblock \emph{IEEE Transactions on Multimedia}, 2021.

\bibitem[Puigcerver et~al.(2024)Puigcerver, Ruiz, Mustafa, and Houlsby]{puigcerver2024softmoe}
Puigcerver, J., Ruiz, C.~R., Mustafa, B., and Houlsby, N.
\newblock From sparse to soft mixtures of experts.
\newblock In \emph{The Twelfth International Conference on Learning Representations}, 2024.

\bibitem[Riquelme et~al.(2021)Riquelme, Puigcerver, Mustafa, Neumann, Jenatton, Susano~Pinto, Keysers, and Houlsby]{riquelme2021scaling}
Riquelme, C., Puigcerver, J., Mustafa, B., Neumann, M., Jenatton, R., Susano~Pinto, A., Keysers, D., and Houlsby, N.
\newblock Scaling vision with sparse mixture of experts.
\newblock \emph{Advances in Neural Information Processing Systems}, 34:\penalty0 8583--8595, 2021.

\bibitem[Shazeer et~al.(2017)Shazeer, Mirhoseini, Maziarz, Davis, Le, Hinton, and Dean]{shazeer2017moe}
Shazeer, N., Mirhoseini, A., Maziarz, K., Davis, A., Le, Q., Hinton, G., and Dean, J.
\newblock Outrageously large neural networks: The sparsely-gated mixture-of-experts layer.
\newblock \emph{arXiv preprint arXiv:1701.06538}, 2017.

\bibitem[Shi et~al.(2016)Shi, Caballero, Husz{\'a}r, Totz, Aitken, Bishop, Rueckert, and Wang]{shi2016real}
Shi, W., Caballero, J., Husz{\'a}r, F., Totz, J., Aitken, A.~P., Bishop, R., Rueckert, D., and Wang, Z.
\newblock Real-time single image and video super-resolution using an efficient sub-pixel convolutional neural network.
\newblock In \emph{Proceedings of the IEEE Conference on Computer Vision and Pattern Recognition}, pp.\  1874--1883, 2016.

\bibitem[Sun et~al.(2022)Sun, Pan, and Tang]{sun2022shufflemixer}
Sun, L., Pan, J., and Tang, J.
\newblock Shufflemixer: An efficient convnet for image super-resolution.
\newblock \emph{Advances in Neural Information Processing Systems}, 35:\penalty0 17314--17326, 2022.

\bibitem[Sun et~al.(2023)Sun, Dong, Tang, and Pan]{sun2023safmn}
Sun, L., Dong, J., Tang, J., and Pan, J.
\newblock Spatially-adaptive feature modulation for efficient image super-resolution.
\newblock In \emph{Proceedings of the IEEE/CVF International Conference on Computer Vision}, 2023.

\bibitem[Vaswani et~al.(2017)Vaswani, Shazeer, Parmar, Uszkoreit, Jones, Gomez, Kaiser, and Polosukhin]{vaswani2017attention}
Vaswani, A., Shazeer, N., Parmar, N., Uszkoreit, J., Jones, L., Gomez, A.~N., Kaiser, {\L}., and Polosukhin, I.
\newblock Attention is all you need.
\newblock \emph{Advances in neural information processing systems}, 30, 2017.

\bibitem[Wang et~al.(2023)Wang, Chen, Ni, Liu, and jinfan]{wang2023omnisr}
Wang, H., Chen, X., Ni, B., Liu, Y., and jinfan, L.
\newblock Omni aggregation networks for lightweight image super-resolution.
\newblock In \emph{Conference on Computer Vision and Pattern Recognition}, 2023.

\bibitem[Wang et~al.(2021)Wang, Xie, Dong, and Shan]{wang2021realesrgan}
Wang, X., Xie, L., Dong, C., and Shan, Y.
\newblock Real-esrgan: Training real-world blind super-resolution with pure synthetic data.
\newblock In \emph{Proceedings of the IEEE/CVF international conference on computer vision}, pp.\  1905--1914, 2021.

\bibitem[Wang et~al.(2022)Wang, Su, Li, Cao, Wang, and Liu]{wang2022ddistill}
Wang, Y., Su, T., Li, Y., Cao, J., Wang, G., and Liu, X.
\newblock Ddistill-sr: Reparameterized dynamic distillation network for lightweight image super-resolution.
\newblock \emph{IEEE Transactions on Multimedia}, 2022.

\bibitem[Yang et~al.(2022)Yang, Li, and Gao]{yang2022focalnet}
Yang, J., Li, C., and Gao, J.
\newblock Focal modulation networks.
\newblock arXiv, 2022.

\bibitem[Yu et~al.(2021)Yu, Wang, Dong, Tang, and Loy]{yu2021path}
Yu, K., Wang, X., Dong, C., Tang, X., and Loy, C.~C.
\newblock Path-restore: Learning network path selection for image restoration.
\newblock \emph{IEEE Transactions on Pattern Analysis and Machine Intelligence}, 44\penalty0 (10):\penalty0 7078--7092, 2021.

\bibitem[Zamir et~al.(2022)Zamir, Arora, Khan, Hayat, Khan, and Yang]{Zamir2021Restormer}
Zamir, S.~W., Arora, A., Khan, S., Hayat, M., Khan, F.~S., and Yang, M.-H.
\newblock Restormer: Efficient transformer for high-resolution image restoration.
\newblock In \emph{CVPR}, 2022.

\bibitem[Zeyde et~al.(2010)Zeyde, Elad, and Protter]{zeyde2010single}
Zeyde, R., Elad, M., and Protter, M.
\newblock On single image scale-up using sparse-representations.
\newblock In \emph{Proceedings of International Conference on Curves and Surfaces}, pp.\  711--730. Springer, 2010.

\bibitem[Zhang et~al.(2021{\natexlab{a}})Zhang, Li, Luo, Ren, Stenger, Liu, Li, and Yang]{zhang2021benchmarking}
Zhang, K., Li, D., Luo, W., Ren, W., Stenger, B., Liu, W., Li, H., and Yang, M.-H.
\newblock Benchmarking ultra-high-definition image super-resolution.
\newblock In \emph{Proceedings of the IEEE/CVF international conference on computer vision}, pp.\  14769--14778, 2021{\natexlab{a}}.

\bibitem[Zhang et~al.(2021{\natexlab{b}})Zhang, Zeng, and Zhang]{zhang2021ecbsr}
Zhang, X., Zeng, H., and Zhang, L.
\newblock Edge-oriented convolution block for real-time super resolution on mobile devices.
\newblock In \emph{Proceedings of the ACM International Conference on Multimedia}, 2021{\natexlab{b}}.

\bibitem[Zhang et~al.(2022)Zhang, Zeng, Guo, and Zhang]{zhang2022elan}
Zhang, X., Zeng, H., Guo, S., and Zhang, L.
\newblock Efficient long-range attention network for image super-resolution.
\newblock In \emph{Proceedings of the European Conference on Computer Vision}, 2022.

\bibitem[Zhang et~al.(2018{\natexlab{a}})Zhang, Li, Li, Wang, Zhong, and Fu]{zhang2018rcan}
Zhang, Y., Li, K., Li, K., Wang, L., Zhong, B., and Fu, Y.
\newblock Image super-resolution using very deep residual channel attention networks.
\newblock In \emph{Proceedings of the European conference on computer vision}, 2018{\natexlab{a}}.

\bibitem[Zhang et~al.(2018{\natexlab{b}})Zhang, Tian, Kong, Zhong, and Fu]{zhang2018residual}
Zhang, Y., Tian, Y., Kong, Y., Zhong, B., and Fu, Y.
\newblock Residual dense network for image super-resolution.
\newblock In \emph{Proceedings of the IEEE Conference on Computer Vision and Pattern Recognition}, 2018{\natexlab{b}}.

\bibitem[Zhao et~al.(2020)Zhao, Kong, He, Qiao, and Dong]{zhao2020pan}
Zhao, H., Kong, X., He, J., Qiao, Y., and Dong, C.
\newblock Efficient image super-resolution using pixel attention.
\newblock In \emph{Proceedings of the European Conference on Computer Vision Workshops}, pp.\  56–72, 2020.

\bibitem[Zhou et~al.(2023)Zhou, Li, Guo, Bai, Cheng, and Hou]{zhou2023srformer}
Zhou, Y., Li, Z., Guo, C.-L., Bai, S., Cheng, M.-M., and Hou, Q.
\newblock Srformer: Permuted self-attention for single image super-resolution.
\newblock \emph{Proceedings of the IEEE/CVF International Conference on Computer Vision}, 2023.

\end{thebibliography}
\bibliographystyle{icml2024}

\appendix
\twocolumn[

\icmltitle{See More Details: Efficient Image Super-Resolution by Experts Mining  \\ -- Appendix --}

\vskip 0.13in
]

\begin{table}[ht]
    \centering
    \footnotesize
    \setlength\tabcolsep{1.5pt}
    \caption{\textit{Implementation Details.}}
    \label{tab:supp:arch_config}
    \begin{tabularx}{\columnwidth}{X*{3}{c}}
    \toprule
     Parameter & SeemoRe-T & SeemoRe-B & SeemoRe-L \\
    \midrule
    Num. RGs & 6  & 8 & 16 \\
    Channel dimension & 36 & 48 & 48 \\
    MLP-Ratio &  \multicolumn{3}{c}{2}\\
    LR dimensionality growth & \multicolumn{3}{c}{exponential}\\
    Num. Experts $\mathcal{E}$ & \multicolumn{3}{c}{3} \\
    Top-$k$ experts &  \multicolumn{3}{c}{1} \\
    SFM kernel size &  \multicolumn{3}{c}{11} \\
    Recursion steps & 2 & 2 & 1 \\
    Training Dataset & \multicolumn{3}{c}{DIV2K + Flickr2K}\\
    Optimizer & \multicolumn{3}{c}{Adam} \\
    Batch size & \multicolumn{3}{c}{32} \\
    Total Num. Iterations & \multicolumn{3}{c}{$500k$} \\
    FFT Loss weight & \multicolumn{3}{c}{0.1} \\
    LR-Rate & \multicolumn{3}{c}{1e.3}\\
    LR-Decay Rate & \multicolumn{3}{c}{0.5} \\
    LR-Decay Milestones & \multicolumn{3}{c}{[$250K$,$400K$,$450K$,$475K$]}\\
    \bottomrule
    \end{tabularx}
\end{table}%

\begin{table}[t]
    \centering
    \footnotesize
    \caption{\textit{Ablation on contribution of components.} GMACS ($\downarrow$) are computed by upscaling to a $1280\times720$ HR image. ($^\ast$) denotes modified configuration from proposed SeemoRe-T model.}
    \label{tab:supp:block_contribution}
    \setlength\tabcolsep{1pt}
    \begin{tabularx}{\columnwidth}{X*{6}{c}}
        \toprule
        Method & RME & SME & Params. & GMACS & Urban100 & Manga109 \\
        \midrule
        Baseline$^\ast$ & - & - & 232K & 52 & 31.73 & 38.63  \\
        \multirow{3}{*}{SeemoRe-T}& $\checkmark$ & - & 249K & 48 & 31.99 &	38.75 \\
         & -  & $\checkmark$ & 238K & 53 & 32.05 &	38.96 \\
         & \cellcolor{red!10}{$\checkmark$} & \cellcolor{red!10}{$\checkmark$} & \cellcolor{red!10}{\textbf{220K}} & \cellcolor{red!10}{\textbf{45}} & \cellcolor{red!10}{\textbf{32.22}} & \cellcolor{red!10}{\textbf{39.01}} \\
        \bottomrule
    \end{tabularx} 
\end{table}

\begin{table*}[t]
    \centering
    \footnotesize
    \setlength\tabcolsep{2pt}   
    \caption{\textit{Comparison to lightweight SR Transformers.}~ Extension of \cref{tab:exp:lightweight_sota}. PSNR (dB $\uparrow$) and SSIM ($\uparrow$) metrics are reported on the Y-channel. GMACS $\downarrow$) are computed by upscaling to a $1280\times720$ HR image.
    }
    \label{tab:supp:lightweight_sota_ext}
    
    \begin{tabularx}{\textwidth}{lX*{14}{c}}
    \toprule
    & \multirow{2}{*}{Method} & \multirow{2}{*}{Params} & \multirow{2}{*}{GMACS} &  
    \multicolumn{2}{c}{SET5} & \multicolumn{2}{c}{SET14} & \multicolumn{2}{c}{BSD100} & \multicolumn{2}{c}{Urban100} & \multicolumn{2}{c}{Manga109} \\
    \cmidrule(lr){5-6} \cmidrule(lr){7-8} \cmidrule(lr){9-10} \cmidrule(lr){11-12} \cmidrule(lr){13-14} 
    & & & &PSNR&SSIM &PSNR&SSIM &PSNR&SSIM &PSNR&SSIM &PSNR&SSIM\\
    
    \midrule

    \multirow{8}*{\rotatebox{90}{$\times 3$}} 
    & Bicubic & - & - & 30.39 & .8682 & 27.55 & .7742 & 27.21 & .7385 & 24.46 & .7349 & 26.95 & .8556\\
    & SwinIR-Light~\cite{liang2021swinir} & 918K & 111  &34.62&.9289&30.54&.8463&29.20&.8082&28.66&.8624&33.98&.9478\\
    & ELAN-Light~\cite{zhang2022elan} & 629K &90 &34.61&.9288&30.55&.8463&29.21&.8081&28.69&.8624&34.00&.9478\\
    & SRFormer-Light~\cite{zhou2023srformer} & 861K &105 & 34.67 &.9296 & 30.57 & .8469 & 29.26 &.8099 & 28.81 & .8655 & 34.19 & .9489\\
    & ESRT~\cite{lu2022esrt} & 770K & 96 &34.42&.9268&30.43&.8433&29.15&.8063&28.46&.8574&33.95&.9455 \\
    & SwinIR-NG~\cite{choi2023n} & 1190K & 114 & 34.64 & .9293 & 30.58 & .8471 & 29.24 & .8090 & 28.75 & .8639 & 34.22 & .9488\\
    & DAT-Light~\cite{chen2023dat} & 629K & 89 & 34.76 & .9299 & 30.63 & .8474 & 29.29 & .8103 & 28.89 & .8666 & 34.55 & .9501 \\
    & SeemoRe-L (\textit{ours}) & 959K & 87 & 34.72 & .9297 & 30.60 & .8469& 29.29 & .8101 & 28.86 & .8653 & 34.53 & .9496\\

    \bottomrule
    \end{tabularx}
\end{table*}%

\begin{table*}[t]
    \centering
    \footnotesize
    \setlength\tabcolsep{3.5pt}   
    \caption{\textit{Ablation on the top-$k$ experts.}~PSNR (dB $\uparrow$) and SSIM ($\uparrow$) metrics are reported on the Y-channel for $\times 2$ upscaling. GMACS ($\downarrow$) and memory consumption (M, $\downarrow$) are computed by upscaling to a $1280\times720$ HR image using a NVIDIA RTX 4090 device.
    }
    \label{tab:supp:top_k}
    
    \begin{tabularx}{\textwidth}{X*{13}{c}}
    \toprule
    \multirow{2}{*}{Method} & \multirow{2}{*}{Params} & \multirow{2}{*}{GMACS} & \multirow{2}{*}{GPU Memory} &
    \multicolumn{2}{c}{SET5} & \multicolumn{2}{c}{SET14} & \multicolumn{2}{c}{BSD100} & \multicolumn{2}{c}{Urban100} & \multicolumn{2}{c}{Manga109} \\
     \cmidrule(lr){5-6} \cmidrule(lr){7-8} \cmidrule(lr){9-10} \cmidrule(lr){11-12} \cmidrule(lr){12-13}
      & & & & PSNR&SSIM &PSNR&SSIM &PSNR&SSIM &PSNR&SSIM &PSNR&SSIM\\
    
    \midrule
    SeemoRe-T \\
    \ \ $k=1$ &  \cellcolor{red!10}{\textbf{220K}} & \cellcolor{red!10}{\textbf{44.83}} & \cellcolor{red!10}{\textbf{10972}} & \cellcolor{red!10}{38.06} & \cellcolor{red!10}{.9608} & \cellcolor{red!10}{33.65} & \cellcolor{red!10}{\textbf{.9186}} & \cellcolor{red!10}{\textbf{32.23}} & \cellcolor{red!10}{\textbf{.9004}} & \cellcolor{red!10}{32.22} & \cellcolor{red!10}{.9286} & \cellcolor{red!10}{39.01} & \cellcolor{red!10}{.9777} \\
    \ \ $k=2$ & 220K & 45.22 & 11233 & 38.09 & .9608 & 33.61 & .9184 & 32.22 & .9003 & 32.23 & .9286 & 39.00 & .9777 \\
    \ \ $k=3$ & 220K & 46.12 & 11494 & \textbf{38.10} & \textbf{.9609} &\textbf{ 33.66} & .9185 & 32.23 & .9004 & \textbf{32.24} & \textbf{.9289} & \textbf{39.08} & \textbf{.9779}\\

    \bottomrule
    \end{tabularx}
\end{table*}%

\begin{table}[t]
    \centering
    \footnotesize
    \setlength\tabcolsep{0.25pt}
    \caption{\textit{Analysis of proposed SEE block.} We have conducted the following experiment by replacing our proposed SEE with the spatial enhancement module, Fused-MBConv in Shufflemixer~\cite{sun2022shufflemixer}, and Conv Block in Conv2Former~\cite{hou2022conv2former} on $\times 2$ scale. }
    \label{tab:supp:different_blocks}
    \begin{tabularx}{\columnwidth}{X*{4}{c}}
    \toprule
    Method & Params. & GMACS & Urban100 & Manga109 \\
    \midrule
    SeemoRe-T \\
    FusedMB~\cite{sun2022shufflemixer} & 304K & 64 & 32.18 & 38.99 \\
    C2F~\cite{hou2022conv2former} & \textbf{220K} & 46 & 32.19 & \textbf{39.04} \\
   SEE (\textit{ours}) & \cellcolor{red!10}{\textbf{220K}} & \cellcolor{red!10}{\textbf{45}} & \cellcolor{red!10}{\textbf{32.22}} & \cellcolor{red!10}{39.01} \\
    \bottomrule
        
    \end{tabularx}
\end{table}%

\begin{table}[t]
    \centering
    \footnotesize
    \setlength\tabcolsep{3pt}
    \caption{\textit{Analysis of MoRE design. } We provide further insights in the design decisions of our SeemoRe framework for $\times 2$ upscaling.}
    \label{tab:supp:more_design}

    \begin{tabularx}{\columnwidth}{lX*{2}{c}}
    \toprule
    Method & Residual $t$ & Urban100 & Manga109 \\
    \midrule
    \multirow{3}{*}{SeemoRe-T} & No aggregation & 32.11 & 38.96 \\
    & Aggregation output & 32.15 & 38.99 \\
    & \cellcolor{red!10}{DConv output} & \cellcolor{red!10}{\textbf{32.22}} & \cellcolor{red!10}{\textbf{39.01}} \\    
    \bottomrule
    \end{tabularx}

\end{table}%

\begin{table}[t]
    \centering
    \footnotesize
        \setlength\tabcolsep{1.5pt}
    \caption{\textit{Optimization function.} SeemoRe-T was trained on DIV2K and Flickr2K. We report PSNR (dB $\uparrow$) on the Y-Channel for $\times 2$ upscaling.}
    \label{tab:supp:loss}
        \begin{tabularx}{\columnwidth}{X*{7}{c}}
                \toprule
                Method & \multicolumn{2}{c}{L1} & \multicolumn{2}{c}{FFT} & BSD100 & Urban100 & Manga109 \\
                \midrule
                \multirow{3}{*}{SeemoRe-T}
                & $\checkmark$ & 1.0 & - & 0.0 & 32.21 & 32.14 & 38.90 \\ 
                & \cellcolor{red!10}{$\checkmark$} & \cellcolor{red!10}{1.0} & \cellcolor{red!10}{$\checkmark$} & \cellcolor{red!10}{0.1} & \cellcolor{red!10}{32.23} & \cellcolor{red!10}{\textbf{32.22}} & \cellcolor{red!10}{39.01} \\
                & $\checkmark$ & 1.0 & $\checkmark$ & 0.2 & 32.22 & 32.16 & \textbf{39.02} \\
                \bottomrule
        \end{tabularx}
\end{table}%

\begin{table}[t]
    \centering
    \footnotesize
    \setlength\tabcolsep{1pt}
    \caption{\textit{Model size.} PSNR (dB $\uparrow$) is reported on the Y-channel. GMACS are computed by upscaling to a $1280\times720$ HR image. N and C denote number of RGs and channel features, respectively.}
    \label{tab:supp:model_sizes}
        \begin{tabularx}{\columnwidth}{X*{6}{c}}
        \toprule
        Method & \multicolumn{2}{c}{Config} & Params. & GMACS & Urban100 & Manga109 \\
        \midrule
        SeemoRe-T &  N:6 & C:36 & 220K & 45 & 32.22  & 39.01 \\
        SeemoRe-B &  N:8 & C:48 & 490K & 101 & 32.52	& 39.30\\
        SeemoRe-L & N:16 & C:48 & 931K & 197 & 32.87 & 39.49 \\
        \bottomrule
    \end{tabularx}
\end{table}

\begin{table}[t]
    \centering
    \footnotesize
    \setlength\tabcolsep{1pt}
    \caption{\textit{Scaling up the numbers of experts.} We analyze the impact of the number of experts on SeemoRe-T's performance.}
    \label{tab:supp:num_exp}
    \begin{tabularx}{\columnwidth}{lX*{5}{c}}
    \toprule
    Scale & Method & $\#\mathcal{E}$ & Growth & Params. & Urban100 & Manga109 \\
    \midrule
    \multirow{3}{*}{\rotatebox{90}{$\times 2$}} & \multirow{3}{*}{SeemoRe-T} & 8 & $2*i + 2$ & 261K & 32.18 & \textbf{39.02} \\
    & & 4 & $2^{i}$ & 231K & 32.21 & \textbf{39.02} \\
    & & \cellcolor{red!10}{3} & \cellcolor{red!10}{$2^{i}$} & \cellcolor{red!10}{\textbf{220K}} & \cellcolor{red!10}{\textbf{32.22}} & \cellcolor{red!10}{39.01} \\
    \bottomrule
    \end{tabularx}
\end{table}%

\section{Further Implementation Details}
\cref{tab:supp:arch_config} outlines the architectural configurations and training settings employed to achieve the reported results in this study. Throughout all our experiments, we maintained a fixed random seed for reproducibility purposes. We based our implementation on the public PyTorch-based \textit{BasicSR}\footnote{\url{https://github.com/XPixelGroup/BasicSR}} framework for architecture development and training. We use \textit{fvcore}\footnote{\url{https://github.com/facebookresearch/fvcore}} Python package for computing GMACS and parameter counts.

\paragraph{Baseline for Architecture Contribution.} Here we provide more details about the baseline method for the ablation in \cref{tab:exp:arch_components,tab:supp:block_contribution}. In the main text, we evaluate SeemoRe-T by sequentially removing the proposed RME and SME blocks, resulting in a plain baseline model with fewer parameters and GMACs. To ensure a fair comparison, we adjust the baseline configuration to match our plain SeemoRe-T model. To ensure roughly equivalent parameter counts and computational complexity, we adopt $5$ RGs with a channel dimensionality of $48$. Within each RG, we integrate simple convolutional operators from our RME submodule without the MoRE module, while the SME module is simplified to a pointwise convolution.

\paragraph{Comparison to lightweight SR Transformers ($\times 3$).}
In \cref{tab:supp:lightweight_sota_ext}, we present the performance of our SeemoRe-L model for $\times 3$ upscaling, extending the results from \cref{tab:exp:lightweight_sota} in the main text. Our SeemoRe-L consistently outperforms other lightweight Transformers and demonstrates only a slightly lower performance compared to DAT-Light~\cite{chen2023dat}.

\section{More Ablations}
\subsection{Architecture Design}
\paragraph{SEE block compared to prior designs.}
The results in \cref{tab:supp:different_blocks} prove that our SEE block design outperforms the FusedMB-Conv block proposed by ShuffleMixer~\cite{sun2022shufflemixer} in terms of reconstruction abilities while maintaining higher efficiency. Moreover, substituting the large-kernel convolution in the Conv2Former~\cite{hou2022conv2former} block with our striped large-kernel variant not only enhances efficiency but also improves the reconstruction capabilities of high-frequency information, as evident from Urban100 results.

\paragraph{MoRE block design.}
Our rationale behind the MoRE design involves the aggregation of valuable contextual information. Similar to prior works~\cite{liu2020residual}, we assign more learning parameters to enhance the high-frequency features while keeping the simple DConv-branch as residual to facilitate the optimization. We further support this rationale with empirical evidence provided in \cref{tab:supp:more_design}. The results show that adding the extended feature to the output of DConv performs better than with and without the aggregation output.

\paragraph{Optimization function.}
In \cref{tab:supp:loss}, we explore the impact of using the L1-Norm in FFT space to compare the model output with high-quality GT images. Compared to utilizing only the traditional L1 loss in RGB space, we observe an average performance improvement of $0.09$ dB on Urban100 and Manga109 datasets while using the combined losses. We acknowledge that only a few previous methods incorporate the same FFT loss \cite{sun2022shufflemixer,sun2023safmn}; however, other efficient image super-resolution methods either employ a more intricate training schedule with multiple stages \cite{liu2020rfdn,kong2022rlfn} or utilize large-scale models for knowledge distillation \cite{wang2022ddistill}.

\paragraph{Scaling the model size.}
In Table \ref{tab:supp:model_sizes}, we detail the architecture, efficiency, and PSNR results across different model sizes on Urban100 and Manga109 datasets. Starting with SeemoRe-T, which has 220K parameters and 45 GMACS, each subsequent complexity stage doubles these figures. Notably, all model stages achieve state-of-the-art performance within their weight classes, with SeemoRe-L matching or even surpassing recent lightweight Transformer-based SR models.

Futhermore, we investigate increasing the number of experts to $8$ as shown in \cref{tab:supp:num_exp} the impact on the overall model performance. The results indicate that increasing the number of experts adds complexity, however it doesn't consistently improve the reconstruction fidelity.
Balancing the low-rank space and the expert count offers to fine-tune the performance trade-off. Though, our emphasis here is on efficiency, we aim to explore more complex designs in future research.

\subsection{Evaluation on Real SR}
We conduct experiments for Real SR ($\times4$) using the Real-ESRGAN~\cite{wang2021realesrgan} degradation model on SeemoRe-T and the current efficient SOTA SR model SAFMN~\cite{sun2023safmn}, see \cref{tab:supp:realsr}.
Both SAMFN and SeemoRe-T are initialized from the $\times 4$ bicubic checkpoints, we reduce the number of iterations on the DF2K\_OST dataset by half (250k) and train only using the L1 loss.
We report the popular NR-IQA metrics (NIQE and BRISQUE) on the commonly used real-world image collection given in SwinIR~\cite{liang2021swinir}. Additionally, we conduct a cross-dataset evaluation using testsets with more realistic degradation of different severity levels (Type I and Type II), as provided by \cite{liang2022efficient}.

\begin{table}[t]
    \centering
    \footnotesize
    \fboxsep0.75pt
    \setlength\tabcolsep{1pt}
    \caption{\textit{Real SR performance.} NIQE and BRISQUE are reported on the real image collection provided by SwinIR~\cite{liang2021swinir}. DIV2K-I and DIV2K-II performance reported as  \colorbox{blue!10}{PSNR}.}
    \label{tab:supp:realsr}
    \begin{tabularx}{\columnwidth}{X*{4}{c}}
    \toprule
    Method & NIQE~($\downarrow$) & BRISQUE~($\downarrow$) & DIV2K-I & DIV2K-II \\
    \midrule
    Bicubic & 7.65 & 58.29 &  \cellcolor{blue!10}{26.30} & \cellcolor{blue!10}{25.71}\\
    SAFMN & 7.19 & 51.39 &  \cellcolor{blue!10}{26.80} &  \cellcolor{blue!10}{26.77} \\
    SeemoRe-T & \textbf{6.53} & \textbf{45.53} & \cellcolor{blue!10}{\textbf{27.07}} & \cellcolor{blue!10}{\textbf{27.01}} \\
    \bottomrule
        
    \end{tabularx}

\end{table}

\section{Future work and limitations}
\begin{figure*}[t]
    \centering
    \scriptsize
    \setlength\tabcolsep{0.75pt}
        \begin{tabularx}{\textwidth}{*{6}{C}}
            \multicolumn{2}{c}{\includegraphics[width=0.33\textwidth, trim=0 20cm 0 0, clip]{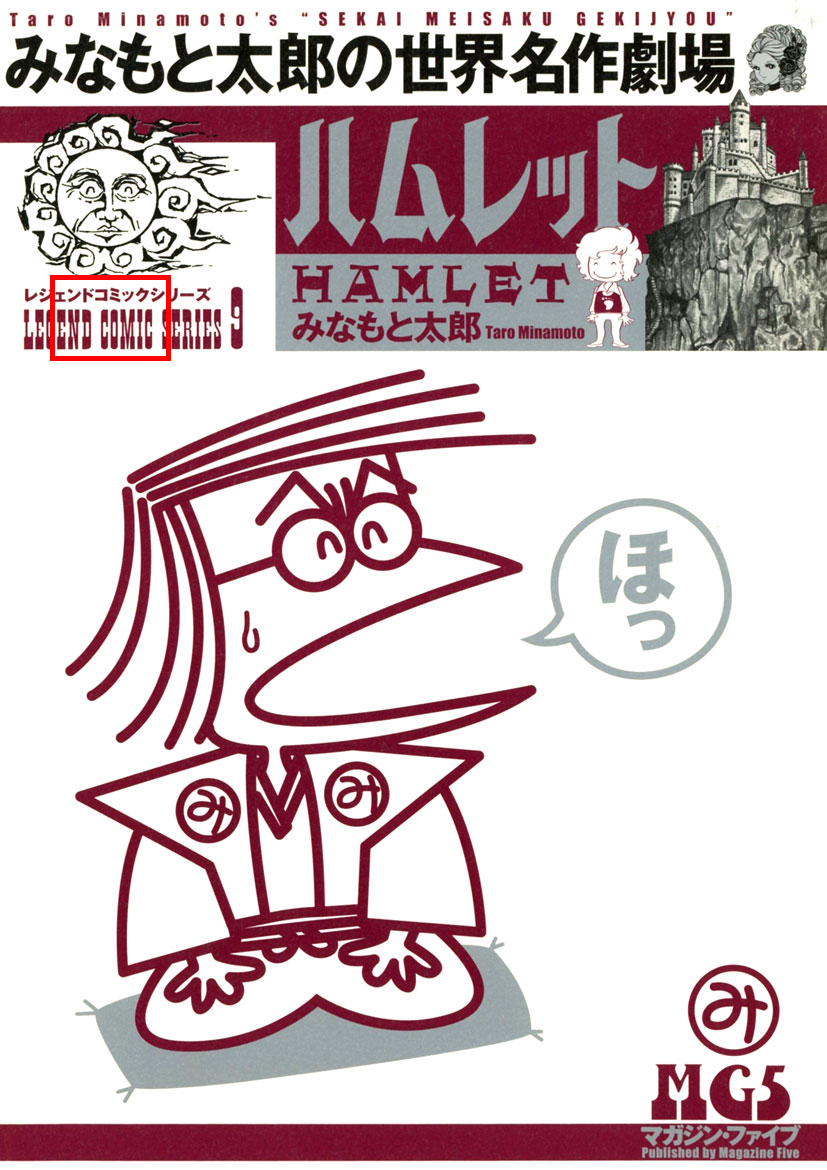}} &
            \multicolumn{2}{c}{\includegraphics[width=0.33\textwidth, trim=0 20cm 0 0, clip]{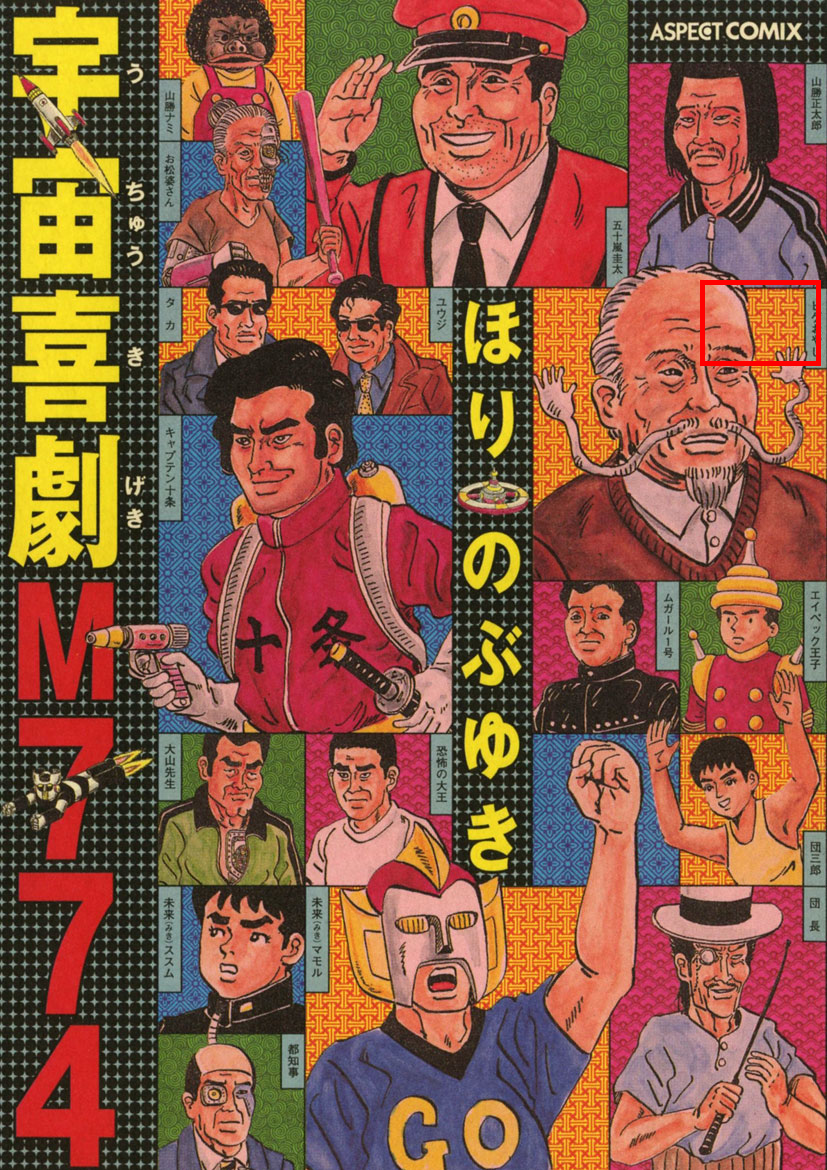}} &
            \multicolumn{2}{c}{\includegraphics[width=0.33\textwidth, trim=0 20cm 0 0, clip]{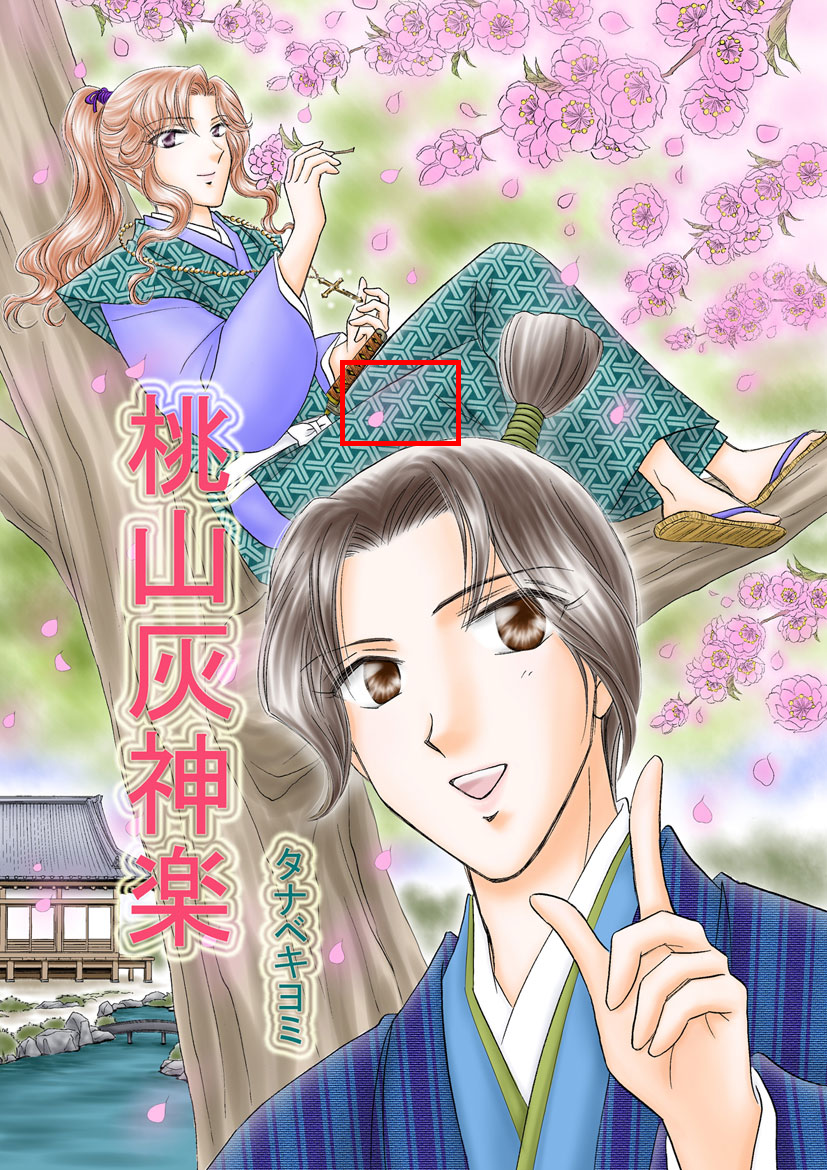}}  \\

            \multicolumn{2}{c}{Manga109: \text{img25} ($\times 4$)} &
            \multicolumn{2}{c}{Manga109: \text{img99} ($\times 4$)} &
            \multicolumn{2}{c}{Manga109: \text{img59} ($\times 4$)} \\

            \includegraphics[width=0.16\textwidth]{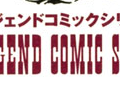} &
            \includegraphics[width=0.16\textwidth]{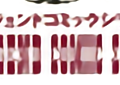} &
            \includegraphics[width=0.16\textwidth]{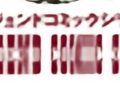} &
            \includegraphics[width=0.16\textwidth]{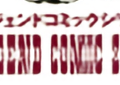} &
            \includegraphics[width=0.16\textwidth]{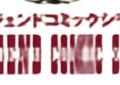} &
            \includegraphics[width=0.16\textwidth]{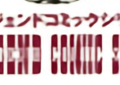} \\

            \includegraphics[width=0.16\textwidth]{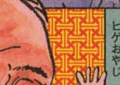} &
            \includegraphics[width=0.16\textwidth]{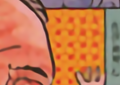} &
            \includegraphics[width=0.16\textwidth]{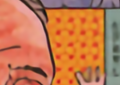} &
            \includegraphics[width=0.16\textwidth]{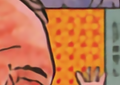} &
            \includegraphics[width=0.16\textwidth]{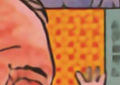} &
            \includegraphics[width=0.16\textwidth]{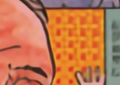} \\

            \includegraphics[width=0.16\textwidth]{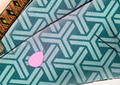} &
            \includegraphics[width=0.16\textwidth]{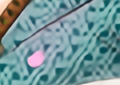} &
            \includegraphics[width=0.16\textwidth]{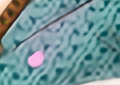} &
            \includegraphics[width=0.16\textwidth]{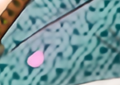} &
            \includegraphics[width=0.16\textwidth]{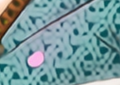} &
            \includegraphics[width=0.16\textwidth]{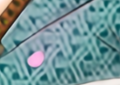} \\
            HR Crop & DDistill-SR & ShuffleMixer & SwinIR-Light &  DAT-Light & SeemoRe-L \\       
        \end{tabularx}

    \caption{Visual comparison of SeemoRe with state-of-the-art methods on challenging cases for $\times 4$ SR from the Manga109 benchmark.}
    \label{fig:supp:visual_manga}
\end{figure*}

\begin{figure*}[t]
    \centering
    \scriptsize
    \setlength\tabcolsep{0.75pt}
        \begin{tabularx}{\textwidth}{*{6}{C}}
            \multicolumn{2}{c}{\includegraphics[width=0.33\textwidth, trim=0 0 0 4.5cm, clip]{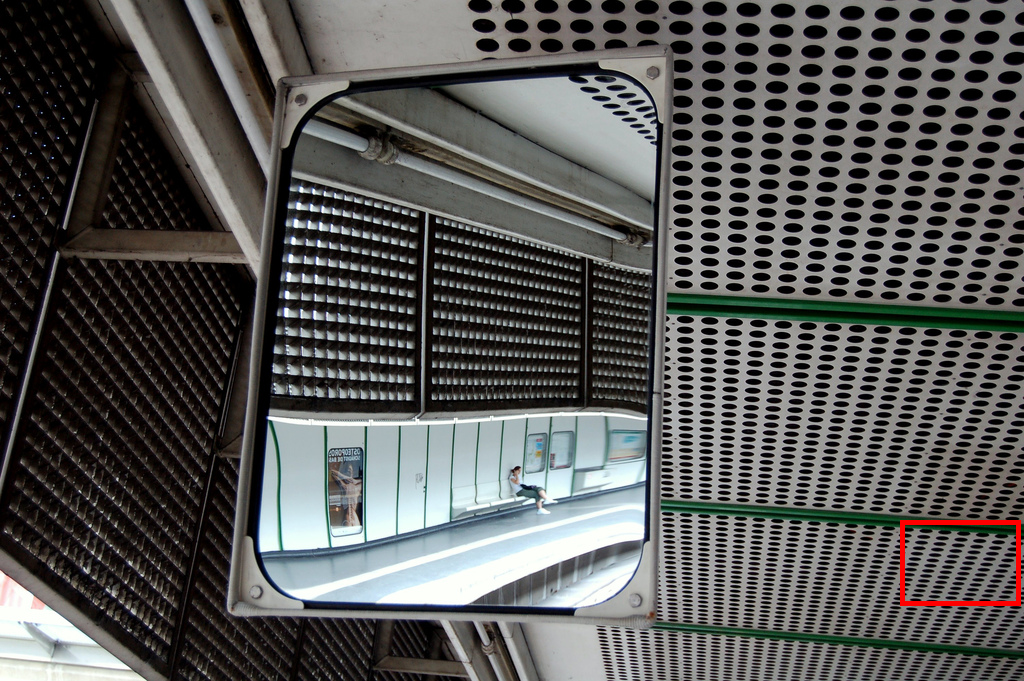}} &
            \multicolumn{2}{c}{\includegraphics[width=0.33\textwidth, trim=0 0 0 23cm, clip]{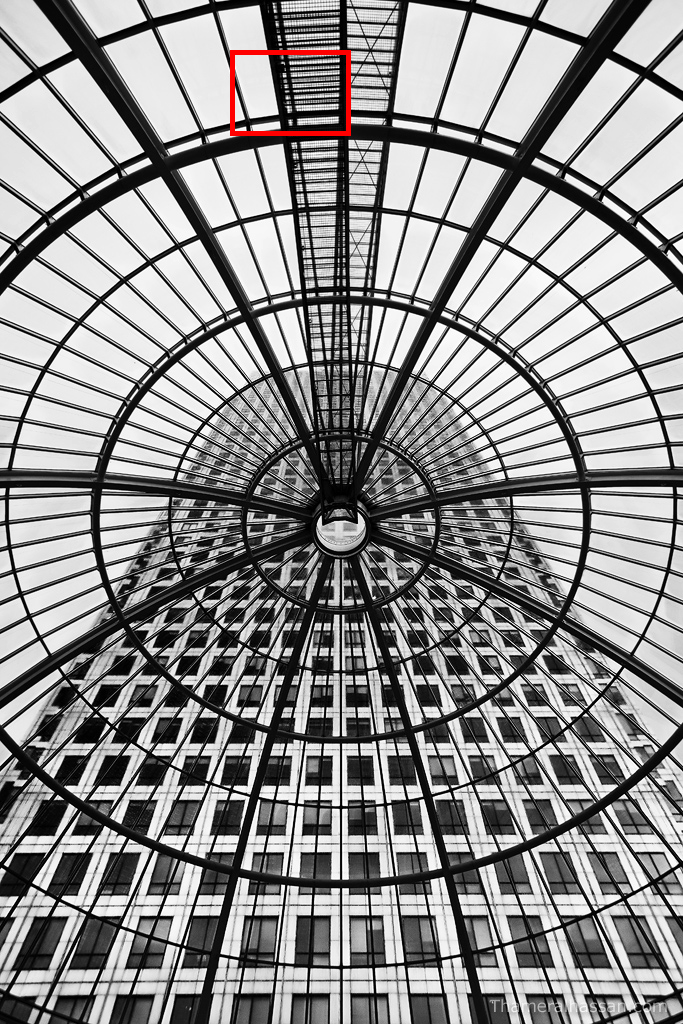}} &
            \multicolumn{2}{c}{\includegraphics[width=0.33\textwidth, trim=0 4.6cm 0 0, clip]{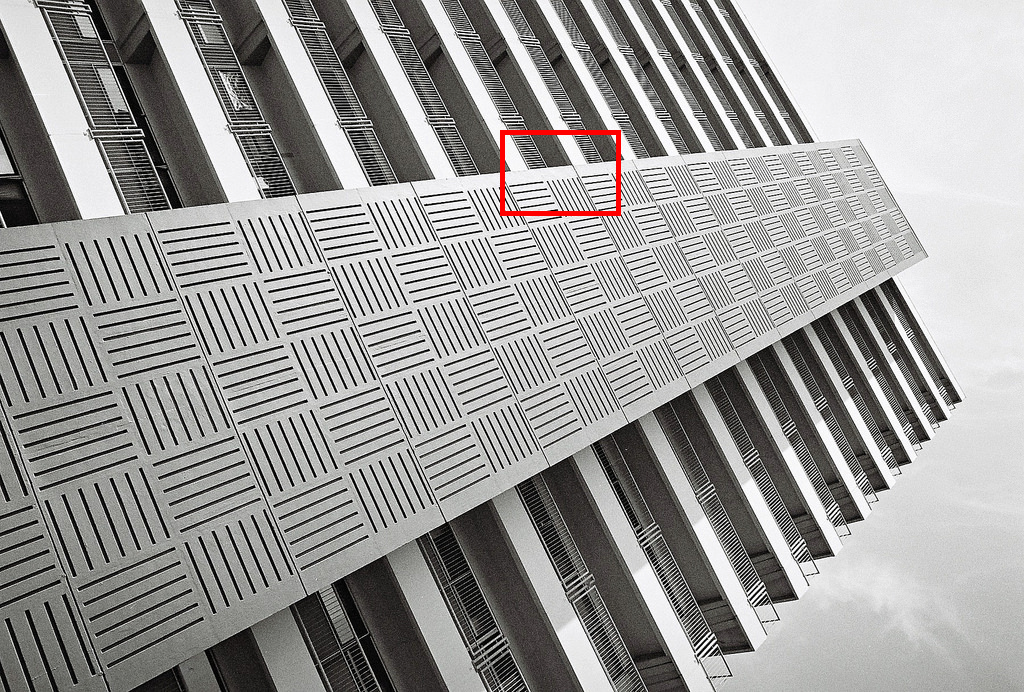}}  \\

            \multicolumn{2}{c}{Urban100: \text{img04} ($\times 4$)} &
            \multicolumn{2}{c}{Urban100: \text{img72} ($\times 4$)} &
            \multicolumn{2}{c}{Urban100: \text{img92} ($\times 4$)} \\

            \includegraphics[width=0.16\textwidth]{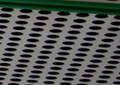} &
            \includegraphics[width=0.16\textwidth]{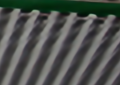} &
            \includegraphics[width=0.16\textwidth]{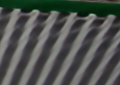} &
            \includegraphics[width=0.16\textwidth]{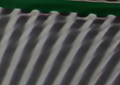} &
            \includegraphics[width=0.16\textwidth]{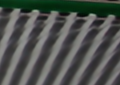} &
            \includegraphics[width=0.16\textwidth]{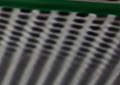} \\

            \includegraphics[width=0.16\textwidth]{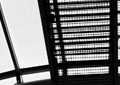} &
            \includegraphics[width=0.16\textwidth]{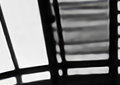} &
            \includegraphics[width=0.16\textwidth]{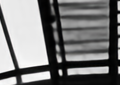} &
            \includegraphics[width=0.16\textwidth]{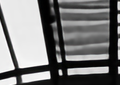} &
            \includegraphics[width=0.16\textwidth]{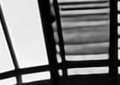} &
            \includegraphics[width=0.16\textwidth]{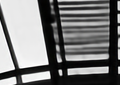} \\

            \includegraphics[width=0.16\textwidth]{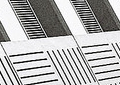} &
            \includegraphics[width=0.16\textwidth]{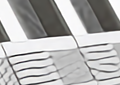} &
            \includegraphics[width=0.16\textwidth]{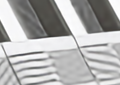} &
            \includegraphics[width=0.16\textwidth]{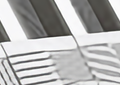} &
            \includegraphics[width=0.16\textwidth]{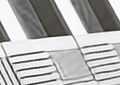} &
            \includegraphics[width=0.16\textwidth]{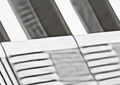} \\
            HR Crop & DDistill-SR & ShuffleMixer & SwinIR-Light &  DAT-Light & SeemoRe-L \\       
        \end{tabularx}

    \caption{Visual comparison of SeemoRe with state-of-the-art methods on challenging cases for $\times 4$ SR from the Urban100 benchmark.}
    \label{fig:supp:visual2_urban100}
\end{figure*}

The proposed approach, employing a mixture of experts for feature modulation, is versatile for tasks with limited input information, such as low-light enhancement and denoising. Additionally, SeemoRe's efficient design makes it a valuable solution for dynamic and resource-intensive environments. Expanding the number of experts in our network's low-rank aspect poses challenges due to rapid feature dimensionality growth. Thus, our approach is currently limited to a small number of experts, contrasting with other fields leveraging larger expert ensembles.
Despite the improving trade-off between efficiency and reconstruction fidelity, as depicted in \cref{fig:supp:visual2_urban100,fig:supp:visual_manga}, our SeemoRe model still contends with blur artifacts. However, similar artifacts can also be observed in Transformer-based super-resolution alternatives, albeit at a higher computational cost (in terms of inference time).
While our model represents a pioneering effort in utilizing a mixture of low-rank experts for super-resolution, significant opportunities for further research exist. For instance, exploring explicit constraints on the features learned by different experts presents intriguing research directions with potential applications across a spectrum of restoration problems. We wish our network serve as a straightforward yet effective baseline, stimulating continued exploration in the field.

\section{Visual Results.}
We provide additional visual comparisons ($\times$4) in \cref{fig:supp:visual_manga} for the Manga109 benchmark and in \cref{fig:supp:visual2_urban100} for the Urban100 benchmark. Our SeemoRe framework consistently produces visually pleasing results, even on artistic images. In contrast to previous methods which exhibit flawed texture and character reconstruction, our proposed approach effectively reconstructs missing details, as illustrated in \cref{fig:supp:visual_manga}, across all exemplary images considered. More concretely, when examining the example image \textit{img25} our SeemoRe network proficiently reconstructs the capital letter ``I" within the text prompt ``COMIC," whereas SwinIR-Light and DAT-L encounter difficulty in producing any readable output. Additionally, in example image \textit{img04} our model significantly outperforms others in reconstructing the pattern with higher fidelity. Moreover, our model's reconstruction of \textit{img92} in \cref{fig:supp:visual2_urban100} demonstrates reduced blurring and more distinct edges, enhancing overall visibility.

\end{document}